# Quasar-galaxy associations $^\star$

P. A. Thomas[1], R. L. Webster[2] and M. J. Drinkwater[3]

[1] Astronomy Centre, MAPS, University of Sussex, Brighton, BN1 9QH
[2] School of Physics, University of Melbourne, Parkville, Victoria 3052, Australia
[3] Anglo-Australian Observatory, Coonabarabran, NSW 2357, Australia

**ABSTRACT**
There is controversy about the measurement of statistical associations between bright quasars and faint, presumably foreground galaxies. We look at the distribution of galaxies around an unbiased sample of 63 bright, moderate redshift quasars using a new statistic based on the separation of the quasar and its nearest neighbour galaxy. We find a significant excess of close neighbours at separations less than about 10 arcsec which we attribute to the magnification by gravitational lensing of quasars which would otherwise be too faint to be included in our sample. About one quarter to one third of the quasars are so affected although the allowed error in this fraction is large.

## 1 INTRODUCTION

In this paper we look at the association between quasars and foreground galaxies. We find a significant excess over that expected from a random distribution which we interpret as being due to gravitational lensing, i.e. many of the quasars in our sample are intrinsically faint but are magnified into the sample by gravitational lensing.

The physical significance of quasar-galaxy associations depends on the redshifts of the quasars and galaxies. In the case of low-redshift ($z \lesssim 0.7$) quasars we might expect to detect associations due to galaxies lying in clusters associated with the quasar. Several instances of low-redshift clustering are reported in the literature (e.g. Yee & Green 1987, Smith & Heckman 1990 and Ellingson, Green & Yee 1991), but in the present study we are not interested in galaxies at the quasar redshift.

The situation for high-redshift ($z \gtrsim 1$) quasars is much less clear. Associations have been detected in recent observations, but the cause has not been unequivocally established. If the galaxies are at lower redshifts than the quasars, there is evidence that these are biased lines-of-sight. The amplification of the background sources may be due either to macrolensing by the galaxy potential, or to microlensing by compact objects associated with the galaxy. Alternatively, the galaxies may be in groups at the quasar redshift. If this is the case, the galaxies must be much brighter than at the present epoch in order to have been detected at such a high redshift.

Until recently, most studies detected a statistically significant excess of galaxies near moderately high redshift quasars (Tyson 1986, Webster et al. 1988, Fugmann 1988, 1989 and Hintzen, Romanishin & Valdes 1991). However, some more recent studies (e.g. Crampton, McClure & Fletcher 1992 and Yee, Filippenko & Tang 1992), claim no statistical excess, though quantitative results have yet to be published. Comparison of different samples is difficult, since different methodologies have been used, and the samples of quasars observed vary markedly in redshift and magnitude. In addition, the algorithm for galaxy detection varies. A discussion of the observational problems is given by Hewett, Harding & Webster (1992). In addition, different groups attribute the excess to different physical reasons. That issue will only be clearly resolved when a substantial number of redshifts are obtained for the associated galaxies.

Because of the hints of gravitational lensing of various degrees of significance in previous work we set out to design an observing programme which would be optimal for observing lensing, if it were present. The choice of quasar sample is described at length in Drinkwater, Webster & Thomas (1993, Paper 1) but the relevant points are reiterated below. We chose to look at bright ($m_V \lesssim 18.5$), moderate redshift ($1 \lesssim z \lesssim 2.5$) quasars: the bright apparent magnitude enhances the fraction of lensed quasars, while the high redshift gives a reasonable cross-section for lensing. It is of crucial importance that the sample be unbiased with respect to the presence of galaxies near quasars: the authors of papers on MgII samples claim the quasar selection was random. To test the possibility that there has been an unintentional favouring of quasars whose spectra show MgII absorption lines, we analysed the subsample of quasars without detected MgII separately. These show a similar associ-ation of quasars and foreground galaxies (see Section 3.2). Preliminary results from essentially the same sample were presented by Drinkwater et al. (1992), but we now apply a more rigorous statistical analysis.

The outline of the paper is as follows. The selection criteria and data analysis are briefly described in Section 2—for a fuller account we refer the reader to Paper 1. In Section 3 we describe the statistics of quasar-galaxy associations. We first discuss the use of the 's-parameter' as a measure of quasar-galaxy separation and then show that it is significantly biased to low values. The number of fields which show associated galaxies is then quantified. Section 4 assesses the validity of our results (we can think of no reasonable explanation other than gravitational lensing) and compares them with previous work. We then go on to look at the properties of individual lensing galaxies and of the population as a whole. The significance for cosmology is discussed. Finally, Section 5 presents a summary of the most important features of our results.

## 2 DATA

Our data consist of R-band (Kron-Cousins) CCD images of bright quasars observed at the Observatoire astronomique du Mont Mégantic, Québec. Most of these were part of the study of MgII absorption systems described in Paper 1, although some objects are excluded from the present sample because we apply stricter selection criteria. We also include additional objects not in Paper 1. Our observing procedure and data reduction are essentially the same as described in Paper 1 with some modifications that we describe in this Section.

### 2.1 Selection Criteria

To measure quasar-galaxy associations we require a sample of bright quasars that is unbiased with respect to the presence of galaxies near the quasars. This condition means that we must exclude 5 objects from the sample in Paper 1 (0738+313, 1049+616, 1127−145, 1148+387 and 1332+552) which were taken from references that only listed detections of MgII absorption and may therefore be biased. We also excluded the quasar 0957+561 since it was added to the Young, Sargent & Boksenberg (1982) compilation because it was known to be gravitationally lensed and therefore was biased towards having a close galaxy (the lens in this case).

To this list we add 16 more quasars. 12 of these are from the Kiso survey (see Wegner & Swanson, 1990b). They were selected by colour and morphological criteria from optical photographic plates and have no bias to the presence of galaxies or absorption. The 4 remaining objects were selected at random from the Hewitt & Burbidge (1989) catalogue (subject to the magnitude and redshift constraints described below) to fill a gap in our observing programme. Two of them were excluded from our statistical analysis as they are below our redshift cutoff (see Section 3.2); the others do not show a particularly close association of quasar and nearest-neighbor galaxy.

Combining the above we have a sample of 80 quasars satisfying the following criteria:

(i)   $m_V \lesssim 18$
(ii)  $z_{qso} \leq 2.5$.
(iii) no bias with respect to nearby galaxies or absorption.
(iv)  Declination $\geq -10$ degrees.
(v)   Right Ascension determined by the allocated observing times.

The full list of quasars observed is given in Table 1.

None of the samples we used were strictly magnitude limited, but all are of bright quasars. The majority of quasar magnitudes in our final sample are in the range $15.5 \leq m_R \leq 18.5$.

### 2.2 Image Catalogues

The CCD images were processed and calibrated using the IRAF[†] image analysis software. The FOCAS (faint object classification and analysis system) package (Valdes 1982) was then used to detect and classify images in each field. We used the DAOPHOT package to detect and measure any images very close to the quasar which were then added into the FOCAS catalogues. The magnitudes of all the objects were then measured with the aperture photometry routines in IRAF. All these procedures are described in more detail in Paper 1, apart from the following modifications which were made to avoid in bias in our catalogues:

(i)   We use a more conservative magnitude limit, now defined as the magnitude of the brightest object with a photometry error $\Delta m > 0.2$ mag in each frame. This results in limits a few tenths of a magnitude brighter than the previous definition (the magnitude of the faintest object with a photometry error $\Delta m < 0.2$ mag). In fact the results are not sensitive to the exact value of the magnitude error chosen. It can be reduced to 0.16 magnitudes before the signal from galaxies associated with quasars begins to decrease.

(ii)  We include only those objects which lie more than 10 pixels from the edge of the field. This is to avoid incompleteness which might otherwise occur for bright objects which extend over the field boundary.

(iii) We have checked the extra objects included from the DAOPHOT analysis to ensure they all have significant detections: all were in fact clearly visible in the raw images.

(iv)  A new set of control stars has been defined using an objective algorithm to avoid any bias. These satisfy the following criteria: distance from the edge $\geq 40$ arcsec, distance from the quasar $\geq 30$ arcsec, and magnitude range $16 \leq m_R \leq 19.0$. If more than one star per field satisfied these criteria, one of them was chosen at random. In some cases no star satisfied these criteria, so only 57 fields have control stars.

Table 2 lists all detected objects within 30 arcsec of each quasar. The effect of the DAOPHOT analysis is to add an extra galaxy close to the quasar in fields 0736−063

---

**Table 1.** Observing details for the quasars.

| (1) QSO | (2) type | (3) RA (1950) | (4) Dec | (5) $z_{qso}$ | (6) ref | (7) date | (8) exp | (9) fwhm | (10) phot | (11) $m_R$ | (12) $m_{lim}$ |
|---|---|---|---|---|---|---|---|---|---|---|---|
| 0710+118 | RX | 7:10:15.38 | 11:51:23.9 | 0.768 | 1 | 15/04/91 | 2100 | 2.00 | p | 16.05 | 21.22 |
| 0736−063 | R | 7:36:30.20 | -6:20:02.9 | 1.914 | 2 | 17/03/91 | 1800 | 2.16 | | 16.73 | 21.52 |
| 0742+318 | R | 7:42:30.75 | 31:50:15.7 | 0.462 | 6 | 30/03/90 | 2400 | 1.61 | | 15.65 | 22.50 |
| 0835+580 | RX | 8:35:10.02 | 58:04:51.8 | 1.534 | 1,3 | 28/03/90 | 2400 | 1.75 | | 16.85 | 21.89 |
| 0836+195 | R | 8:36:15.00 | 19:32:24.6 | 1.691 | 1,3 | 17/03/91 | 1902 | 2.23 | | 17.23 | 22.36 |
| 0838+355 | C | 8:38:07.00 | 35:55:21.0 | 2.53 | 10 | 18/03/91 | 1800 | 2.38 | | 16.46 | 21.78 |
| 0843+136 | R | 8:43:01.35 | 13:39:57.4 | 1.875 | 1,3 | 26/03/90 | 2500 | 1.89 | | 17.10 | 21.88 |
| 0848+163 | C | 8:48:53.70 | 16:23:39.8 | 1.925 | 1,2,3,7 | 13/04/91 | 1800 | 2.16 | | 17.36 | 22.18 |
| 0854+191 | CR | 8:54:36.54 | 19:07:00.5 | 1.896 | 1,3,7 | 13/04/91 | 1800 | 2.06 | | 17.60 | 22.21 |
| 0856+170 | CR | 8:56:04.09 | 17:03:09.1 | 1.449 | 1,3 | 29/03/90 | 2400 | 2.27 | | 17.78 | 22.15 |
| 0856+189 | C | 8:56:37.45 | 18:55:32.2 | 1.286 | 1 | 13/04/91 | 1800 | 1.81 | | 17.67 | 22.00 |
| 0935+414 | C | 9:35:49.00 | 41:41:55.0 | 1.94 | 4 | 18/03/91 | 1200 | 2.16 | d | 15.89 | 21.82 |
| 0952+179 | R | 9:52:11.83 | 17:57:44.9 | 1.472 | 1,3 | 18/03/91 | 1819 | 2.33 | | 17.20 | 21.93 |
| 0955+326 | RX | 9:55:25.44 | 32:38:23.0 | 0.533 | 1 | 14/04/91 | 1800 | 2.02 | | 15.96 | 22.06 |
| 0958+551 | C | 9:58:08.05 | 55:09:05.8 | 1.751 | 1,3,7 | 28/03/90 | 2400 | 1.80 | d | 15.97 | 22.47 |
| 1017+280 | C | 10:17:06.00 | 28:00:60.0 | 1.928 | 2,7 | 17/03/91 | 1800 | 2.21 | | 15.73 | 22.16 |
| 1038+064 | RX | 10:38:40.87 | 6:25:58.6 | 1.27 | 1 | 22/03/90 | 1800 | 2.51 | | 16.26 | 21.53 |
| 1054−034 | OXR | 10:54:10.35 | -3:24:38.7 | 2.115 | 2,7 | 26/03/90 | 2400 | 1.95 | | 17.82 | 22.27 |
| 1055−045 | OR | 10:55:40.18 | -4:34:08.5 | 1.428 | 1,3 | 30/03/90 | 2300 | 1.75 | d | 17.27 | 22.46 |
| 1103−006 | R | 11:03:58.07 | -0:36:37.7 | 0.426 | 1 | 14/04/91 | 2400 | 2.38 | | 16.35 | 21.93 |
| 1104+167 | R | 11:04:36.66 | 16:44:17.1 | 0.634 | 1 | 15/04/91 | 1800 | 1.72 | p | 16.08 | 21.83 |
| 1115+080 | C | 11:15:41.50 | 8:02:24.0 | 1.718 | 2,7 | 18/03/91 | 1800 | 2.28 | | 15.89 | 22.40 |
| 1126+101 | R | 11:26:38.70 | 10:08:32.0 | 1.515 | 1,3 | 29/03/90 | 2400 | 2.61 | | 17.07 | 22.02 |
| 1138+040 | C | 11:38:42.40 | 4:03:38.0 | 1.877 | 7 | 18/03/91 | 1800 | 2.53 | | 16.86 | 22.20 |
| 1146+372 | C | 11:46:44.00 | 37:25:09.0 | 0.80 | 10 | 14/04/91 | 1800 | 1.97 | | 17.55 | 22.54 |
| 1148−001 | R | 11:48:10.17 | -0:07:13.1 | 1.982 | 2,7 | 18/03/91 | 1800 | 2.53 | | 16.99 | 22.12 |
| 1209+107 | O | 12:09:08.40 | 10:46:58.0 | 2.191 | 2 | 30/03/90 | 2331 | 1.50 | | 17.63 | 22.22 |
| 1211+334 | R | 12:11:32.60 | 33:26:18.0 | 1.598 | 1,3 | 30/03/90 | 2400 | 1.53 | d | 17.32 | 22.32 |
| 1215+113 | R | 12:15:53.31 | 11:21:44.6 | 1.396 | 1 | 14/04/91 | 1800 | 2.26 | | 16.69 | 22.27 |
| 1225+317 | RX | 12:25:55.94 | 31:45:12.6 | 2.219 | 2 | 17/03/91 | 1800 | 2.16 | d | 15.52 | 22.49 |
| 1227+292 | C | 12:27:32.00 | 29:20:30.0 | 0.74 | 10 | 13/04/91 | 1800 | 1.70 | d | 16.87 | 22.37 |
| 1228+078 | O | 12:28:01.62 | 7:49:39.2 | 1.813 | 2 | 17/03/91 | 1800 | 2.27 | | 17.33 | 22.12 |
| 1231+292 | C | 12:31:27.00 | 29:24:20.0 | 2.00 | 10 | 14/04/91 | 1800 | 1.84 | | 16.66 | 22.26 |
| 1237+280 | C | 12:37:27.00 | 28:04:36.0 | 1.83 | 9 | 13/04/91 | 1800 | 1.76 | p | 18.15 | 22.55 |
| 1246−057 | O | 12:46:38.80 | -5:42:58.3 | 2.236 | 2 | 18/03/91 | 1800 | 2.59 | d | 16.08 | 22.01 |
| 1246+377 | C | 12:46:28.74 | 37:46:49.7 | 1.242 | 1 | 14/04/91 | 1800 | 2.35 | d | 16.89 | 22.31 |
| 1256+357 | CXR | 12:56:07.84 | 35:44:53.7 | 1.882 | 1,3 | 30/03/90 | 2400 | 1.71 | d | 17.60 | 22.58 |
| 1257+346 | CR | 12:57:26.68 | 34:39:31.4 | 1.375 | 1 | 14/04/91 | 1800 | 2.56 | | 16.38 | 22.02 |
| 1258+283 | C | 12:58:36.00 | 28:35:51.0 | 1.36 | 10 | 15/04/91 | 1800 | 1.28 | p | 16.77 | 22.24 |
| 1258+286 | C | 12:58:23.80 | 28:39:28.0 | 1.922 | 1,3 | 18/03/91 | 1800 | 2.67 | | 17.29 | 21.98 |
| 1303+308 | C | 13:03:32.03 | 30:48:55.2 | 1.759 | 1 | 14/04/91 | 1800 | 2.62 | d | 17.24 | 22.33 |
| 1308+182 | R | 13:08:29.47 | 18:15:33.8 | 1.677 | 1,3 | 30/03/90 | 2400 | 1.93 | | 18.55 | 22.62 |
| 1309−056 | OX | 13:09:00.74 | -5:36:43.4 | 2.224 | 2 | 18/03/91 | 1917 | 2.46 | | 17.28 | 21.95 |
| 1309+340 | C | 13:09:16.86 | 34:02:44.5 | 1.035 | 1 | 17/03/91 | 1800 | 1.78 | | 17.33 | 22.61 |
| 1317+277 | C | 13:17:34.24 | 27:43:52.0 | 1.022 | 1 | 15/04/91 | 1800 | 1.91 | dp | 16.16 | 21.95 |

**Table 1.** – continued

| (1) QSO | (2) type | (3) RA (1950) | (4) Dec | (5) $z_{qso}$ | (6) ref | (7) date | (8) exp | (9) fwhm | (10) phot | (11) $m_R$ | (12) $m_{lim}$ |
|---|---|---|---|---|---|---|---|---|---|---|---|
| 1318+29a | CX | 13:18:53.65 | 29:03:30.3 | 1.703 | 1,3 | 18/03/91 | 1800 | 2.24 | | 17.13 | 22.01 |
| 1318+29b | C | 13:18:54.78 | 29:03:00.6 | 0.549 | 1 | 18/03/91 | 1800 | 2.24 | | 16.30 | 22.01 |
| 1329+412 | C | 13:29:29.90 | 41:17:23.0 | 1.935 | 7 | 26/03/90 | 2400 | 1.69 | | 17.03 | 21.91 |
| 1331+170 | RX | 13:31:10.10 | 17:04:26.0 | 2.081 | 2,7 | 18/03/91 | 1800 | 2.47 | | 16.46 | 22.23 |
| 1346−036 | O | 13:46:08.25 | −3:38:30.5 | 2.344 | 2 | 17/03/91 | 1800 | 1.95 | | 16.86 | 22.58 |
| 1354+195 | R | 13:54:42.08 | 19:33:43.9 | 0.72 | 1 | 26/03/90 | 2400 | 1.98 | | 16.31 | 22.40 |
| 1416+067 | RX | 14:16:38.78 | 6:42:20.9 | 1.436 | 1,3 | 17/03/91 | 1800 | 1.93 | | 16.44 | 22.58 |
| 1416+159 | R | 14:16:27.67 | 15:54:52.1 | 1.472 | 1 | 14/04/91 | 1800 | 2.42 | | 17.77 | 22.23 |
| 1418+255 | C | 14:18:52.00 | 25:52:02.0 | 1.05 | 10 | 18/03/91 | 1800 | 2.59 | | 16.13 | 21.90 |
| 1421+122 | R | 14:21:04.69 | 12:13:26.7 | 1.611 | 2 | 17/03/91 | 1800 | 1.85 | | 17.94 | 22.40 |
| 1421+330 | C | 14:21:17.52 | 33:05:55.5 | 1.904 | 1,3 | 30/03/90 | 2400 | 1.82 | d | 15.98 | 22.02 |
| 1435+638 | R | 14:35:37.20 | 63:49:36.0 | 2.068 | 7 | 22/03/90 | 1800 | 2.05 | | 16.11 | 21.59 |
| 1511+103 | R | 15:11:04.56 | 10:22:15.2 | 1.546 | 1,3 | 25/06/89 | 1500 | 2.65 | | 16.79 | 21.04 |
| 1512+370 | RX | 15:12:46.87 | 37:01:55.2 | 0.371 | 6 | 01/07/89 | 900 | 1.92 | p | 16.20 | 21.32 |
| 1517+239 | C | 15:17:08.19 | 23:56:52.6 | 1.898 | 1,2,3,7 | 14/04/91 | 1800 | 2.69 | | 17.48 | 21.99 |
| 1556+335 | RX | 15:56:59.43 | 33:31:47.4 | 1.646 | 1,3 | 01/07/89 | 1500 | 1.89 | p | 17.20 | 22.10 |
| 1559+173 | R | 15:59:04.63 | 17:22:36.5 | 1.944 | 1,3 | 26/03/90 | 2400 | 2.06 | | 18.21 | 22.30 |
| 1602−001 | R | 16:02:22.11 | -0:11:00.2 | 1.625 | 2 | 26/03/90 | 2400 | 2.15 | | 17.19 | 22.32 |
| 1621+391 | C | 16:21:24.00 | 39:16:27.0 | 1.98 | 9 | 13/04/91 | 1500 | 2.00 | p | 19.53 | 22.30 |
| 1621+412 | C | 16:21:40.00 | 41:23:56.0 | 1.62 | 10 | 17/03/91 | 1800 | 1.72 | | 16.71 | 22.44 |
| 1630+374 | C | 16:30:15.50 | 37:44:08.0 | 1.47 | 5 | 14/04/91 | 1800 | 2.83 | d | 15.85 | 21.91 |
| 1631+393 | C | 16:31:19.00 | 39:30:42.0 | 1.00 | 9 | 18/03/91 | 1800 | 2.09 | | 17.56 | 22.36 |
| 1634+176 | R | 16:34:02.73 | 17:41:10.1 | 1.897 | 1,3 | 01/07/89 | 1500 | 1.76 | | 18.57 | 21.13 |
| 1634+706 | CX | 16:34:51.70 | 70:37:37.0 | 1.334 | 8 | 26/06/89 | 1376 | 1.86 | d | 13.67 | 21.73 |
| 1641+399 | RX | 16:41:17.62 | 39:54:10.7 | 0.595 | 1,6 | 01/07/89 | 1500 | 1.76 | p | 17.44 | 22.09 |
| 1704+608 | RX | 17:04:03.47 | 60:48:31.1 | 0.371 | 6 | 25/06/89 | 1500 | 2.43 | | 14.90 | 21.14 |
| 1715+535 | C | 17:15:30.70 | 53:31:24.0 | 1.929 | 7 | 26/06/89 | 1500 | 1.72 | | 15.40 | 21.77 |
| 1756+237 | RX | 17:56:56.50 | 23:43:55.0 | 1.721 | 1,3 | 25/06/89 | 1500 | 2.49 | | 16.21 | 21.13 |
| 1828+487 | RX | 18:28:13.55 | 48:42:40.4 | 0.692 | 1 | 14/04/91 | 1800 | 2.92 | | 16.59 | 21.74 |
| 1830+285 | R | 18:30:52.40 | 28:31:16.6 | 0.594 | 8 | 01/07/89 | 1500 | 1.65 | | 18.12 | 21.74 |
| 1901+319 | R | 19:01:02.34 | 31:55:15.0 | 0.635 | 8 | 30/06/89 | 1500 | 2.33 | | 16.34 | 21.34 |
| 2120+168 | RX | 21:20:25.53 | 16:51:46.4 | 1.805 | 1,3 | 26/06/89 | 1500 | 2.22 | | 17.38 | 21.69 |
| 2145+067 | RX | 21:45:36.11 | 6:43:41.2 | 0.99 | 1,7 | 30/06/89 | 900 | 2.84 | | 15.39 | 20.94 |
| 2156+297 | R | 21:56:27.71 | 29:44:47.0 | 1.753 | 8 | 30/06/89 | 1500 | 2.48 | | 19.36 | 21.60 |
| 2230+114 | RX | 22:30:07.84 | 11:28:22.8 | 1.037 | 7 | 17/12/90 | 2400 | 2.08 | | 16.63 | 22.58 |

Notes to Table 1

1: quasar name, 2: discovery technique (C; color, O; spectral survey, R; radio, X; X-ray), 3,4: R.A. and Dec. (1950.0), 5: quasar redshift, (1-6 from Hewitt & Burbidge 1989), 6: reference (see Table 2), 7: date observed (UT), 8: exposure time (s), 9: FWHM seeing (arcsec), 10: photometry flag (blank; photometric, p; not photometric, d; no DAOPHOT fitting was possible), 11: quasar R magnitude, 12: limiting R magnitude.

References for Table 1

1: Weymann *et. al.*(1979), 2: Young *et. al.*(1982), 3: Foltz *et. al.*(1986), 4: Wegner & McMahan (1987), 5: Wegner & McMahan (1988), 6: Tytler *et. al.*(1987), 7: Sargent *et. al.*(1988), 8: Hewitt & Burbidge (1989), 9: Wegner & Swanson (1990a), 10: Wegner & Swanson (1990b).

(quasar-galaxy separation 3.9 arcsec), 1416+067 (8.2 arcsec), 1634+176 (2.3 arcsec) and 1756+237 (4.1 arcsec). In addition several galaxies were added to field 0742+318 and one to field 1704+608, however these quasars are of low redshift and are subsequently omitted from the statistics (see Section 3.2). Similarly stars close to the quasar are added in several fields and some merged pairs are split into two. The list given here differs slightly from that given in Table 3 of Paper 1 because of the different selection criteria used.

We also note that the quasar 1115+080 is multiply imaged by gravitational lensing. The DAOPHOT analysis revealed two main components to the quasar image and we have used the mean coordinate of the two as the quasar position. We have also checked that the quasar was not included in the original reference because it was known to be lensed. Note that we do not, in fact, detect the lensing galaxy (Christian, Crabtree & Waddell, 1987) which is hidden under the quasar image.

## 3 STATISTICS OF THE QUASAR-GALAXY ASSOCIATIONS

In this section we will first test the hypothesis that the quasars and galaxies in our fields are independently distributed in position, which we will find to be strongly rejected. Then we will consider estimates of the number of quasars with an associated 'excess' galaxy.

### 3.1 The $s$-parameter

The fundamental parameter which we choose to work with is the separation of the quasar and its nearest-neighbour galaxy (nng). This has the advantages that its statistical properties are easy to compute and it weights each quasar field equally—for example a quasar which is located behind a rich cluster of galaxies will not swamp the statistics.

In all tests for associations the critical unknown is the surface density of galaxies. This is likely to change depending upon the observational technique, exposure time, seeing and position on the sky. It is therefore difficult to combine data from different fields or different surveys. For this reason we normalise the background counts to the number of detected galaxies *in the same field*. This removes all external bias which might affect the significance of the associations. The possibility of a bias introduced by our observational technique or data reduction is discussed in Section 3.2, below.

Consider a circle of radius $r$ about the quasar. The probability that any particular galaxy does not lie within the circle is $1 - \pi r^2/A$ where $A$ is the total area in which galaxies could be detected. If we assume that the galaxies are scattered at random across the field then the probability that there is no galaxy within the circle is $(1 - \pi r^2/A)^N$ where $N$ is the total number of galaxies. Hence the probability that the quasar and the nng have separation greater than $r$ is

$$s = 1 - \left(1 - \frac{\pi r^2}{A}\right)^N.$$

We use the value of $s$ corresponding to the quasar-nng separation as our fundamental random variable. If the galaxies really are distributed at random then the $s$ ought to be uniformly distributed in the interval [0,1]. In practice we find a significant bias towards small separations.

There are several points worthy of note:

(i) This definition of $s$ differs slightly from that in Paper 1. There we assumed a Poisson distribution of galaxies of *known* surface density $\sigma = N/A$. This overlooks the fact that the measured value of $\sigma$ will be scattered about the true value. We then had

$$s = 1 - \exp(-\sigma \pi r^2).$$

It is easy to see that these two definitions agree in the limit that $N$ becomes very large. In practice the difference between the two is small.

(ii) There are areas in each field in which galaxies would escape detection. These include regions close to the quasar and other bright stars. The former will be the most important and once again will tend to reinforce our conclusions. In principle one should divide the useful area of the chip into two regions at separations less than and greater than $r$. Then $s$ would be

$$s = 1 - \left(1 - \frac{A(< r)}{A}\right)^N,$$

The effect of this change is to enhance the significance of the associations as will be discussed in Sections 3.2 and 3.3, below.

(iii) If the galaxies are clustered in the field then $s$ will not be uniform. In effect we will have a lower $N$ than otherwise because if one galaxy is located at a distance of greater than $r$ from the quasar then there will be a higher probability that the others will also. This raises the expected value of $s$ and strengthens our conclusions. In fact clustering of galaxies in our data appears to be negligible (see Section 4.2) although there is significant field-to-field variation in the number counts. This is to be expected as a typical cluster will fill the frame, even at $z = 0.5$.

Table 3 shows the value of the s-parameter for both galaxies and stars which are brighter than the magnitude limit in each field, as described in Section 2.2. Also shown are the number of detected objects, the magnitude of the nng, and its separation from the quasar. We have also looked at the results using a variety of brighter magnitude limits. As the magnitude cut is lowered so the number of objects decreases and the s-parameter increases until such time as the nng is itself excluded from the sample. This will be discussed in detail in the following Sections.

### 3.2 Testing for associations

We adopt as our null hypothesis that the galaxies are randomly distributed in each field. Then $s$ should be drawn from a uniform distribution

$$U = \begin{cases} 1, & 0 < s < 1 \\ 0, & \text{otherwise} \end{cases}.$$

Quasar-galaxy associations will tend to lower $s$ to small values (typically $s \lesssim 0.1$). We use two statistics to test for deviation from U. The first of these is the Kolmogorov-Smirnov

**Table 2.** Images within 30 arcsec of the quasars.

| ΔRA | ΔDec | sep | cls | $m_R$ | $\sigma_{m_R}$ | ΔRA | ΔDec | sep | cls | $m_R$ | $\sigma_{m_R}$ |
|---|---|---|---|---|---|---|---|---|---|---|---|
| \multicolumn{12}{c}{QSO 0710+118, $z = 0.768$, seeing=2.0 arcsec, magnitude limit=21.22} | | | | | | | | | | | |
| -25.5 | -13.3 | 28.7 | g | 19.23 | 0.02 | -3.9 | -6.8 | 7.9 | s | 18.50 | 0.01 |
| -21.9 | 10.0 | 24.0 | s | 19.58 | 0.03 | 5.2 | -0.9 | 5.2 | s | 15.93 | 0.00 |
| -17.1 | 23.6 | 29.1 | fg | 21.84 | 0.24 | 6.2 | 6.7 | 9.1 | s | 15.53 | 0.01 |
| -14.8 | 21.5 | 26.1 | fs | 21.42 | 0.16 | 6.9 | 7.9 | 10.5 | s | 14.83 | 0.01 |
| -4.8 | -10.7 | 11.8 | s | 18.79 | 0.02 | 17.5 | 16.3 | 23.9 | s | 19.01 | 0.02 |
| -3.9 | -19.1 | 19.5 | fs | 21.31 | 0.16 | 18.4 | 0.3 | 18.4 | fs | 20.91 | 0.02 |
| 0.0 | 0.0 | 0.0 | Q | 16.05 | 0.02 | 19.0 | -6.6 | 20.1 | s | 18.83 | 0.02 |
| 1.2 | 16.2 | 16.3 | s | 19.52 | 0.04 | 28.2 | -1.0 | 28.2 | s | 20.13 | 0.05 |
| \multicolumn{12}{c}{QSO 0736−063, $z = 1.914$, seeing=2.2 arcsec, magnitude limit=21.52} | | | | | | | | | | | |
| -23.5 | 0.4 | 23.5 | fg | 21.30 | 0.14 | 0.0 | 0.1 | 0.1 | Q | 16.73 | 0.00 |
| -18.9 | -11.3 | 22.1 | s | 20.54 | 0.06 | 2.9 | 29.5 | 29.6 | fs | 21.29 | 0.14 |
| -17.5 | 10.3 | 20.3 | fs | 21.88 | 0.23 | 10.2 | -6.8 | 12.3 | s | 20.43 | 0.07 |
| -10.9 | 21.3 | 23.9 | s | 17.10 | 0.00 | 10.3 | 15.9 | 18.9 | fs | 21.54 | 0.15 |
| -10.5 | 5.1 | 11.6 | g | 20.87 | 0.09 | 15.0 | -8.3 | 17.2 | fs | 21.40 | 0.15 |
| -6.5 | -22.3 | 23.2 | s | 20.85 | 0.08 | 17.0 | 1.0 | 17.0 | fg | 21.52 | 0.17 |
| -5.7 | 2.1 | 6.1 | s | 17.89 | 0.01 | 19.3 | -10.2 | 21.8 | s | 20.22 | 0.05 |
| -1.6 | -3.5 | 3.9 | g | 20.10 | 0.04 | 19.9 | 0.9 | 19.9 | s | 21.14 | 0.11 |
| \multicolumn{12}{c}{QSO 0742+318, $z = 0.462$, seeing=1.6 arcsec, magnitude limit=22.50} | | | | | | | | | | | |
| -23.5 | -5.2 | 24.0 | s | 16.25 | 0.00 | -0.1 | 0.0 | 0.1 | Q | 15.65 | 0.00 |
| -18.2 | -9.6 | 20.6 | fg | 23.05 | 0.16 | 6.9 | 23.6 | 24.6 | s | 18.61 | 0.01 |
| -13.9 | -10.6 | 17.5 | s | 15.98 | 0.00 | 7.9 | 10.3 | 13.0 | s | 17.20 | 0.00 |
| -9.6 | -10.7 | 14.4 | fs | 21.57 | 0.04 | 8.9 | -17.6 | 19.7 | fg | 22.36 | 0.17 |
| -8.4 | 18.4 | 20.2 | fs | 22.05 | 0.12 | 16.0 | -20.5 | 26.0 | g | 20.61 | 0.04 |
| -3.7 | 3.7 | 5.3 | fg | 22.51 | 0.13 | 18.4 | 15.0 | 23.8 | fs | 21.97 | 0.12 |
| \multicolumn{12}{c}{QSO 0835+580, $z = 1.534$, seeing=1.8 arcsec, magnitude limit=21.89} | | | | | | | | | | | |
| -29.8 | 0.6 | 29.8 | fg | 22.68 | 0.40 | 3.3 | 6.0 | 6.9 | g | 21.58 | 0.15 |
| -15.3 | 20.7 | 25.7 | fs | 23.05 | 0.56 | 4.5 | 21.7 | 22.2 | g | 21.65 | 0.15 |
| -0.7 | 9.9 | 10.0 | fs | 22.45 | 0.35 | 9.1 | 0.5 | 9.1 | g | 21.06 | 0.10 |
| 0.0 | 0.0 | 0.0 | Q | 16.85 | 0.00 | 11.5 | -18.2 | 21.5 | s | 20.02 | 0.04 |
| 2.8 | -19.9 | 20.1 | s | 17.84 | 0.01 | 16.9 | -22.1 | 27.8 | fg | 22.61 | 0.39 |
| \multicolumn{12}{c}{QSO 0836+195, $z = 1.691$, seeing=2.2 arcsec, magnitude limit=22.36} | | | | | | | | | | | |
| 0.0 | 0.0 | 0.0 | Q | 17.23 | 0.00 | 10.4 | 1.1 | 10.5 | fg | 22.06 | 0.16 |
| 0.1 | -16.3 | 16.3 | s | 19.40 | 0.01 | | | | | | |
| \multicolumn{12}{c}{QSO 0838+355, $z = 2.530$, seeing=2.4 arcsec, magnitude limit=21.78} | | | | | | | | | | | |
| -18.7 | 8.7 | 20.6 | fg | 21.61 | 0.16 | 0.6 | -13.3 | 13.3 | fs | 21.40 | 0.14 |
| 0.0 | 0.0 | 0.0 | Q | 16.46 | 0.00 | | | | | | |
| \multicolumn{12}{c}{QSO 0843+136, $z = 1.875$, seeing=1.9 arcsec, magnitude limit=21.88} | | | | | | | | | | | |
| -18.9 | 5.1 | 19.6 | g | 19.64 | 0.03 | 7.5 | -18.4 | 19.9 | fg | 21.96 | 0.23 |
| -6.2 | 23.4 | 24.3 | g | 20.73 | 0.07 | 11.1 | 11.7 | 16.1 | g | 20.17 | 0.04 |
| 0.0 | 0.0 | 0.0 | Q | 17.10 | 0.00 | 15.1 | -5.2 | 15.9 | s | 15.79 | 0.00 |
| 1.9 | -25.9 | 25.9 | fg | 20.69 | 0.06 | 29.6 | -1.3 | 29.6 | fs | 21.75 | 0.18 |
| 4.1 | -26.5 | 26.8 | g | 20.34 | 0.05 | | | | | | |

**Table 2.** – continued

| ΔRA | ΔDec | sep | cls | $m_R$ | $\sigma_{m_R}$ | ΔRA | ΔDec | sep | cls | $m_R$ | $\sigma_{m_R}$ |
|---|---|---|---|---|---|---|---|---|---|---|---|
| \multicolumn{12}{c}{QSO 0848+163, $z = 1.925$, seeing=2.2 arcsec, magnitude limit=22.18} |
| -13.3 | -12.2 | 18.0 | s | 18.16 | 0.01 | 2.9 | -14.9 | 15.2 | g | 19.21 | 0.01 |
| -6.3 | -27.2 | 27.9 | fg | 21.79 | 0.16 | 9.3 | 19.4 | 21.5 | fg | 22.16 | 0.19 |
| 0.0 | 0.0 | 0.0 | Q | 17.36 | 0.00 | | | | | | |
| \multicolumn{12}{c}{QSO 0854+191, $z = 1.896$, seeing=2.1 arcsec, magnitude limit=22.21} |
| 0.1 | 0.0 | 0.1 | Q | 17.60 | 0.00 | 12.7 | 0.6 | 12.8 | fg | 22.21 | 0.20 |
| 8.4 | 23.6 | 25.0 | fs | 22.57 | 0.29 | 13.0 | -2.8 | 13.3 | s | 21.26 | 0.09 |
| 9.2 | 9.4 | 13.1 | fs | 22.69 | 0.32 | | | | | | |
| \multicolumn{12}{c}{QSO 0856+170, $z = 1.449$, seeing=2.3 arcsec, magnitude limit=22.15} |
| -10.4 | 0.8 | 10.4 | g | 21.45 | 0.15 | 10.2 | 10.7 | 14.8 | s | 15.97 | 0.00 |
| -9.6 | 11.7 | 15.1 | s | 18.53 | 0.01 | 13.0 | -14.0 | 19.1 | s | 16.23 | 0.00 |
| 0.0 | 0.0 | 0.0 | Q | 17.78 | 0.01 | | | | | | |
| \multicolumn{12}{c}{QSO 0856+189, $z = 1.286$, seeing=1.8 arcsec, magnitude limit=22.00} |
| 0.0 | 0.0 | 0.0 | Q | 17.67 | 0.00 | 9.4 | 5.6 | 10.9 | fs | 22.80 | 0.41 |
| \multicolumn{12}{c}{QSO 0935+414, $z = 1.940$, seeing=2.2 arcsec, magnitude limit=21.82} |
| -18.5 | 17.2 | 25.3 | g | 19.42 | 0.02 | 0.0 | 0.0 | 0.0 | Q | 15.89 | 0.00 |
| -12.2 | 7.1 | 14.1 | s | 21.31 | 0.13 | 15.2 | -3.7 | 15.7 | s | 20.58 | 0.06 |
| -9.5 | -3.3 | 10.0 | s | 23.10 | 0.67 | | | | | | |
| \multicolumn{12}{c}{QSO 0952+179, $z = 1.472$, seeing=2.3 arcsec, magnitude limit=21.93} |
| -11.6 | -19.4 | 22.6 | g | 21.00 | 0.09 | 21.5 | 10.7 | 24.0 | s | 21.13 | 0.09 |
| -3.5 | -26.1 | 26.3 | s | 21.10 | 0.09 | 25.8 | -2.9 | 26.0 | g | 20.79 | 0.07 |
| 0.0 | 0.0 | 0.0 | Q | 17.20 | 0.00 | 26.3 | -0.3 | 26.3 | s | 20.71 | 0.07 |
| 9.4 | 9.4 | 13.3 | s | 19.51 | 0.02 | | | | | | |
| \multicolumn{12}{c}{QSO 0955+326, $z = 0.533$, seeing=2.0 arcsec, magnitude limit=22.06} |
| -25.5 | -4.9 | 26.0 | s | 21.95 | 0.19 | 18.9 | 18.7 | 26.6 | g | 21.85 | 0.17 |
| -6.6 | 10.3 | 12.2 | s | 22.25 | 0.24 | 22.8 | 13.6 | 26.5 | s | 22.06 | 0.24 |
| 0.0 | 0.0 | 0.0 | Q | 15.96 | 0.00 | 25.5 | 14.2 | 29.2 | s | 21.77 | 0.16 |
| 13.6 | -19.1 | 23.5 | g | 22.11 | 0.22 | | | | | | |
| \multicolumn{12}{c}{QSO 0958+551, $z = 1.751$, seeing=1.8 arcsec, magnitude limit=22.47} |
| -23.8 | -9.8 | 25.7 | fs | 22.90 | 0.29 | -8.7 | 21.8 | 23.4 | g | 19.30 | 0.01 |
| -20.7 | -1.0 | 20.8 | fs | 22.47 | 0.20 | 0.0 | 0.0 | 0.0 | Q | 15.97 | 0.00 |
| -18.6 | -23.0 | 29.6 | fs | 23.81 | 0.62 | 9.8 | -14.9 | 17.8 | fs | 21.90 | 0.11 |
| -17.8 | -13.3 | 22.2 | fs | 23.37 | 0.43 | | | | | | |
| \multicolumn{12}{c}{QSO 1017+280, $z = 1.928$, seeing=2.2 arcsec, magnitude limit=22.16} |
| -2.0 | 22.8 | 22.9 | fs | 22.21 | 0.21 | 5.8 | -18.4 | 19.3 | g | 21.16 | 0.09 |
| 0.0 | 0.1 | 0.1 | Q | 15.73 | 0.00 | 6.1 | 25.9 | 26.7 | s | 20.63 | 0.04 |
| 0.6 | -18.1 | 18.1 | g | 20.59 | 0.05 | | | | | | |
| \multicolumn{12}{c}{QSO 1038+064, $z = 1.270$, seeing=2.5 arcsec, magnitude limit=21.53} |
| -8.1 | 7.9 | 11.3 | s | 16.50 | 0.00 | 9.6 | 1.7 | 9.7 | g | 21.03 | 0.16 |
| 0.0 | 0.0 | 0.0 | Q | 16.26 | 0.00 | 13.9 | 15.0 | 20.5 | g | 19.22 | 0.03 |
| \multicolumn{12}{c}{QSO 1054−034, $z = 2.115$, seeing=2.0 arcsec, magnitude limit=22.27} |
| -17.1 | -3.0 | 17.3 | g | 21.16 | 0.08 | 0.0 | 0.0 | 0.0 | Q | 17.82 | 0.00 |
| -7.2 | 25.1 | 26.1 | fs | 23.20 | 0.49 | 18.9 | 17.1 | 25.5 | fs | 22.70 | 0.32 |
| -4.8 | 3.0 | 5.7 | g | 19.93 | 0.03 | 20.4 | -0.2 | 20.4 | g | 22.22 | 0.18 |

**Table 2.** – continued

| ΔRA | ΔDec | sep | cls | $m_R$ | $\sigma_{m_R}$ | ΔRA | ΔDec | sep | cls | $m_R$ | $\sigma_{m_R}$ |
|---|---|---|---|---|---|---|---|---|---|---|---|
| | | | QSO 1055−045, $z = 1.428$, seeing=1.8 arcsec, magnitude limit=22.46 | | | | | | | | |
| -2.0 | 19.4 | 19.5 | s | 21.00 | 0.05 | 15.0 | 12.2 | 19.3 | s | 20.93 | 0.05 |
| 0.0 | 0.0 | 0.0 | Q | 17.27 | 0.00 | 19.6 | -1.6 | 19.6 | s | 18.35 | 0.01 |
| 9.5 | -24.2 | 26.0 | s | 20.78 | 0.04 | 22.1 | -5.7 | 22.9 | s | 20.73 | 0.04 |
| | | | QSO 1103−006, $z = 0.426$, seeing=2.4 arcsec, magnitude limit=21.93 | | | | | | | | |
| -26.3 | -8.5 | 27.6 | g | 21.15 | 0.09 | 9.9 | 10.5 | 14.4 | s | 17.36 | 0.00 |
| -16.6 | 10.0 | 19.4 | s | 18.65 | 0.01 | 10.1 | -28.1 | 29.8 | g | 19.97 | 0.03 |
| -14.2 | -4.5 | 14.9 | g | 19.71 | 0.03 | 16.0 | -13.8 | 21.1 | g | 20.67 | 0.06 |
| 0.0 | 0.0 | 0.0 | Q | 16.35 | 0.00 | | | | | | |
| | | | QSO 1104+167, $z = 0.634$, seeing=1.7 arcsec, magnitude limit=21.83 | | | | | | | | |
| -26.2 | -12.5 | 29.0 | s | 13.74 | SAT | -0.1 | 0.0 | 0.1 | Q | 16.08 | 0.00 |
| -14.1 | -0.5 | 14.1 | fg | 21.51 | 0.14 | 3.9 | 16.9 | 17.3 | s | 20.78 | 0.08 |
| -13.6 | 19.5 | 23.8 | fs | 21.73 | 0.16 | 6.5 | 13.6 | 15.1 | g | 21.02 | 0.10 |
| -12.7 | -22.6 | 25.9 | fs | 21.50 | 0.13 | 16.6 | -22.0 | 27.6 | s | 20.92 | 0.08 |
| | | | QSO 1115+080, $z = 1.718$, seeing=2.3 arcsec, magnitude limit=22.40 | | | | | | | | |
| -20.8 | -10.8 | 23.4 | g | 19.34 | 0.02 | 5.6 | -21.8 | 22.5 | s | 19.00 | 0.01 |
| -14.8 | -11.7 | 18.9 | g | 20.85 | 0.06 | 16.3 | 9.2 | 18.7 | g | 21.63 | 0.11 |
| -12.3 | -0.9 | 12.3 | g | 20.33 | 0.04 | 24.5 | -0.4 | 24.5 | g | 21.67 | 0.11 |
| 0.0 | 0.0 | 0.0 | Q | 15.89 | 0.00 | | | | | | |
| | | | QSO 1126+101, $z = 1.515$, seeing=2.6 arcsec, magnitude limit=22.02 | | | | | | | | |
| 0.0 | 0.0 | 0.0 | Q | 17.07 | 0.00 | 5.0 | 8.9 | 10.2 | g | 20.68 | 0.06 |
| 4.2 | 6.4 | 7.7 | fg | 20.82 | 0.07 | 15.3 | -1.5 | 15.4 | fs | 23.81 | 1.04 |
| | | | QSO 1138+040, $z = 1.877$, seeing=2.5 arcsec, magnitude limit=22.20 | | | | | | | | |
| 0.0 | 0.1 | 0.1 | Q | 16.86 | 0.00 | 12.8 | -11.4 | 17.2 | g | 21.08 | 0.07 |
| 2.4 | -12.9 | 13.1 | s | 16.19 | 0.00 | | | | | | |
| | | | QSO 1146+372, $z = 0.800$, seeing=2.0 arcsec, magnitude limit=22.54 | | | | | | | | |
| 0.0 | 0.0 | 0.0 | Q | 17.55 | 0.00 | 1.6 | -10.7 | 10.8 | g | 21.67 | 0.12 |
| | | | QSO 1148−001, $z = 1.982$, seeing=2.5 arcsec, magnitude limit=22.12 | | | | | | | | |
| -29.1 | 4.4 | 29.5 | g | 21.01 | 0.07 | 0.0 | 0.0 | 0.0 | Q | 16.99 | 0.00 |
| -11.2 | 9.9 | 14.9 | s | 20.60 | 0.05 | 5.9 | 1.2 | 6.0 | fs | 22.45 | 0.32 |
| -3.7 | -12.6 | 13.2 | s | 21.15 | 0.08 | 27.3 | -9.1 | 28.7 | fs | 21.84 | 0.16 |
| | | | QSO 1209+107, $z = 2.191$, seeing=1.5 arcsec, magnitude limit=22.22 | | | | | | | | |
| -19.9 | -1.2 | 19.9 | s | 20.78 | 0.05 | 0.0 | 0.0 | 0.0 | Q | 17.63 | 0.00 |
| -8.5 | 17.4 | 19.4 | s | 21.15 | 0.08 | 5.1 | 5.1 | 7.2 | g | 21.43 | 0.10 |
| -5.4 | 1.1 | 5.5 | s | 20.62 | 0.05 | | | | | | |
| | | | QSO 1211+334, $z = 1.598$, seeing=1.5 arcsec, magnitude limit=22.32 | | | | | | | | |
| -20.5 | 7.9 | 22.0 | g | 21.47 | 0.09 | 10.9 | -27.0 | 29.1 | fg | 22.34 | 0.20 |
| -9.2 | 2.7 | 9.6 | fs | 22.37 | 0.19 | 17.8 | 4.0 | 18.2 | s | 18.03 | 0.00 |
| 0.0 | 0.0 | 0.0 | Q | 17.32 | 0.00 | 26.5 | -9.7 | 28.2 | fs | 22.68 | 0.28 |
| 8.7 | 5.0 | 10.1 | fs | 22.79 | 0.32 | | | | | | |
| | | | QSO 1215+113, $z = 1.396$, seeing=2.3 arcsec, magnitude limit=22.27 | | | | | | | | |
| -20.6 | -3.5 | 20.9 | s | 17.19 | 0.00 | -4.0 | 19.8 | 20.2 | s | 21.66 | 0.14 |
| -11.0 | -4.5 | 11.9 | fs | 21.94 | 0.18 | 0.0 | 0.0 | 0.0 | Q | 16.69 | 0.00 |

**Table 2.** – continued

| ΔRA | ΔDec | sep | cls | $m_R$ | $\sigma_{m_R}$ | ΔRA | ΔDec | sep | cls | $m_R$ | $\sigma_{m_R}$ |
|---|---|---|---|---|---|---|---|---|---|---|---|
| | | | QSO 1225+317, $z = 2.219$, seeing=2.2 arcsec, magnitude limit=22.49 | | | | | | | | |
| -8.0 | 1.3 | 8.1 | fs | 21.56 | 0.11 | 4.4 | 14.6 | 15.3 | s | 18.89 | 0.01 |
| 0.0 | 0.0 | 0.0 | Q | 15.52 | 0.00 | 24.0 | 6.7 | 24.9 | fs | 21.88 | 0.12 |
| | | | QSO 1227+292, $z = 0.740$, seeing=1.7 arcsec, magnitude limit=22.37 | | | | | | | | |
| -4.4 | -4.0 | 6.0 | g | 21.93 | 0.18 | 15.1 | -8.6 | 17.4 | g | 19.77 | 0.02 |
| 0.0 | 0.1 | 0.1 | Q | 16.87 | 0.00 | | | | | | |
| | | | QSO 1228+078, $z = 1.813$, seeing=2.3 arcsec, magnitude limit=22.12 | | | | | | | | |
| -18.1 | -7.1 | 19.5 | s | 19.81 | 0.03 | 10.4 | 3.9 | 11.1 | fs | 22.21 | 0.22 |
| -8.6 | 14.5 | 16.9 | g | 20.99 | 0.06 | 10.9 | -14.3 | 18.0 | s | 20.49 | 0.05 |
| 0.0 | 0.0 | 0.0 | Q | 17.33 | 0.00 | | | | | | |
| | | | QSO 1231+292, $z = 2.000$, seeing=1.8 arcsec, magnitude limit=22.26 | | | | | | | | |
| -29.8 | -1.9 | 29.9 | s | 19.08 | 0.01 | -0.5 | -15.7 | 15.7 | fs | 22.46 | 0.23 |
| -10.3 | -28.1 | 29.9 | g | 19.13 | 0.01 | 0.0 | 0.0 | 0.0 | Q | 16.66 | 0.00 |
| | | | QSO 1237+280, $z = 1.830$, seeing=1.8 arcsec, magnitude limit=22.55 | | | | | | | | |
| 0.1 | 0.0 | 0.1 | Q | 18.15 | 0.01 | 16.5 | 21.0 | 26.7 | s | 18.08 | 0.01 |
| | | | QSO 1246+377, $z = 1.242$, seeing=2.3 arcsec, magnitude limit=22.31 | | | | | | | | |
| -28.2 | 2.3 | 28.2 | fs | 22.88 | 0.36 | 6.1 | -17.0 | 18.0 | fs | 22.43 | 0.26 |
| 0.0 | 0.0 | 0.0 | Q | 16.89 | 0.00 | 20.3 | 20.4 | 28.8 | fs | 21.95 | 0.15 |
| | | | QSO 1246−057, $z = 2.236$, seeing=2.6 arcsec, magnitude limit=22.01 | | | | | | | | |
| -19.7 | 18.1 | 26.7 | fs | 21.95 | 0.20 | 1.5 | -23.6 | 23.7 | fs | 21.84 | 0.17 |
| 0.0 | 0.0 | 0.0 | Q | 16.08 | 0.00 | 12.2 | 11.7 | 16.9 | s | 19.95 | 0.03 |
| | | | QSO 1256+357, $z = 1.882$, seeing=1.7 arcsec, magnitude limit=22.58 | | | | | | | | |
| -27.7 | 4.7 | 28.1 | g | 20.34 | 0.03 | 5.0 | 5.5 | 7.4 | g | 22.06 | 0.12 |
| -8.2 | -17.8 | 19.6 | g | 21.60 | 0.09 | 15.1 | -17.0 | 22.7 | s | 20.98 | 0.05 |
| 0.0 | 0.0 | 0.0 | Q | 17.60 | 0.00 | | | | | | |
| | | | QSO 1257+346, $z = 1.375$, seeing=2.6 arcsec, magnitude limit=22.02 | | | | | | | | |
| 0.0 | 0.0 | 0.0 | Q | 16.38 | 0.00 | 14.7 | 16.5 | 22.1 | s | 21.19 | 0.08 |
| 1.8 | 17.3 | 17.4 | g | 21.32 | 0.09 | | | | | | |
| | | | QSO 1258+286, $z = 1.922$, seeing=2.7 arcsec, magnitude limit=21.98 | | | | | | | | |
| -14.9 | 20.6 | 25.4 | fg | 21.50 | 0.11 | 7.4 | -15.7 | 17.4 | s | 15.95 | 0.00 |
| -6.4 | -27.6 | 28.4 | g | 20.33 | 0.04 | 10.9 | -3.1 | 11.3 | fg | 22.24 | 0.22 |
| 0.0 | 0.0 | 0.0 | Q | 17.29 | 0.00 | 13.9 | 18.3 | 23.0 | s | 19.05 | 0.01 |
| 4.0 | 24.5 | 24.8 | g | 21.09 | 0.08 | 15.9 | -16.4 | 22.8 | fs | 21.98 | 0.21 |
| | | | QSO 1258+283, $z = 1.360$, seeing=1.3 arcsec, magnitude limit=22.24 | | | | | | | | |
| -26.5 | -2.4 | 26.6 | fg | 21.61 | 0.12 | 6.7 | 26.4 | 27.2 | g | 19.93 | 0.03 |
| 0.0 | 0.0 | 0.0 | Q | 16.77 | 0.00 | | | | | | |
| | | | QSO 1303+308, $z = 1.759$, seeing=2.6 arcsec, magnitude limit=22.33 | | | | | | | | |
| -23.1 | 12.6 | 26.3 | fs | 21.94 | 0.14 | 13.3 | 14.8 | 19.9 | g | 20.63 | 0.04 |
| -13.8 | -16.5 | 21.5 | g | 19.59 | 0.02 | 14.3 | 19.2 | 23.9 | fs | 22.62 | 0.29 |
| 0.0 | 0.0 | 0.0 | Q | 17.24 | 0.00 | | | | | | |

**Table 2.** – continued

| ΔRA | ΔDec | sep | cls | $m_R$ | $\sigma_{m_R}$ | ΔRA | ΔDec | sep | cls | $m_R$ | $\sigma_{m_R}$ |
|---|---|---|---|---|---|---|---|---|---|---|---|
| \multicolumn{12}{c}{QSO 1308+182, $z = 1.677$, seeing=1.9 arcsec, magnitude limit=22.62} |
| -22.7 | -10.5 | 25.0 | fs | 22.28 | 0.15 | -1.0 | 14.2 | 14.2 | s | 22.05 | 0.11 |
| -15.3 | -24.3 | 28.7 | fs | 22.75 | 0.23 | 0.0 | 0.0 | 0.1 | Q | 18.55 | 0.01 |
| -12.0 | -12.5 | 17.4 | s | 22.25 | 0.15 | 3.8 | 2.3 | 4.5 | fg | 22.88 | 0.28 |
| -4.7 | 3.1 | 5.6 | s | 21.29 | 0.06 | 24.8 | 16.0 | 29.5 | g | 21.06 | 0.05 |
| \multicolumn{12}{c}{QSO 1309−056, $z = 2.224$, seeing=2.5 arcsec, magnitude limit=21.95} |
| -1.9 | 16.1 | 16.2 | fg | 21.33 | 0.10 | 22.5 | 1.8 | 22.6 | s | 21.15 | 0.08 |
| 0.0 | 0.0 | 0.0 | Q | 17.28 | 0.00 | | | | | | |
| \multicolumn{12}{c}{QSO 1309+340, $z = 1.035$, seeing=1.8 arcsec, magnitude limit=22.61} |
| -21.5 | 18.2 | 28.1 | s | 20.49 | 0.03 | 0.6 | -13.5 | 13.5 | g | 18.78 | 0.01 |
| 0.0 | 0.0 | 0.0 | Q | 17.33 | 0.00 | 2.3 | -6.6 | 7.0 | g | 21.27 | 0.07 |
| \multicolumn{12}{c}{QSO 1317+277, $z = 1.022$, seeing=1.9 arcsec, magnitude limit=21.95} |
| -6.7 | 4.4 | 8.1 | fs | 21.86 | 0.19 | 13.3 | -3.5 | 13.8 | fs | 22.20 | 0.27 |
| 0.0 | 0.1 | 0.1 | Q | 16.13 | 0.00 | 21.9 | 5.6 | 22.6 | fs | 23.01 | 0.57 |
| \multicolumn{12}{c}{QSO 1318+29a, $z = 1.703$, seeing=2.2 arcsec, magnitude limit=22.01} |
| 0.0 | 0.0 | 0.0 | Q | 17.13 | 0.00 | | | | | | |
| \multicolumn{12}{c}{QSO 1318+29b, $z = 0.549$, seeing=2.2 arcsec, magnitude limit=22.01} |
| -12.3 | -19.8 | 23.3 | fg | 22.81 | 0.37 | 0.0 | 0.0 | 0.0 | Q | 16.30 | 0.00 |
| -3.6 | -23.4 | 23.7 | fs | 21.75 | 0.14 | 0.3 | -5.6 | 5.6 | fs | 22.62 | 0.38 |
| \multicolumn{12}{c}{QSO 1329+412, $z = 1.935$, seeing=1.7 arcsec, magnitude limit=21.91} |
| -10.6 | 15.9 | 19.1 | s | 16.99 | 0.00 | 0.9 | -1.4 | 1.7 | s | 21.61 | 0.17 |
| 0.0 | 0.1 | 0.1 | Q | 17.03 | 0.00 | | | | | | |
| \multicolumn{12}{c}{QSO 1331+170, $z = 2.081$, seeing=2.5 arcsec, magnitude limit=22.23} |
| 0.0 | 0.0 | 0.0 | Q | 16.46 | 0.00 | 26.3 | -7.9 | 27.5 | g | 20.96 | 0.07 |
| 17.7 | -9.8 | 20.2 | fs | 21.82 | 0.13 | | | | | | |
| \multicolumn{12}{c}{QSO 1346−036, $z = 2.344$, seeing=2.0 arcsec, magnitude limit=22.58} |
| -18.9 | 22.1 | 29.1 | s | 21.77 | 0.10 | 6.5 | -7.9 | 10.2 | s | 21.50 | 0.08 |
| -10.8 | 17.5 | 20.5 | fs | 21.91 | 0.11 | 10.4 | 1.0 | 10.5 | fs | 22.11 | 0.15 |
| -2.0 | 9.0 | 9.2 | s | 18.86 | 0.01 | 12.7 | -23.8 | 27.0 | fs | 21.72 | 0.09 |
| 0.0 | 0.1 | 0.1 | Q | 16.86 | 0.00 | 18.3 | 16.1 | 24.4 | s | 21.78 | 0.10 |
| 1.0 | 25.0 | 25.0 | g | 21.73 | 0.11 | | | | | | |
| \multicolumn{12}{c}{QSO 1354+195, $z = 0.720$, seeing=2.0 arcsec, magnitude limit=22.33} |
| -21.4 | -11.8 | 24.4 | g | 21.15 | 0.08 | 1.4 | 7.6 | 7.7 | g | 21.60 | 0.13 |
| -16.3 | -22.7 | 28.0 | fs | 22.05 | 0.17 | 3.2 | -6.3 | 7.1 | fg | 21.73 | 0.15 |
| -11.0 | 17.0 | 20.2 | g | 21.32 | 0.08 | 9.8 | -11.8 | 15.4 | fs | 22.97 | 0.40 |
| 0.0 | 0.1 | 0.1 | Q | 16.31 | 0.00 | | | | | | |
| \multicolumn{12}{c}{QSO 1416+159, $z = 1.472$, seeing=2.4 arcsec, magnitude limit=22.23} |
| -22.2 | 3.8 | 22.5 | g | 20.95 | 0.06 | 0.0 | 0.0 | 0.0 | Q | 17.77 | 0.00 |
| -19.9 | -6.1 | 20.8 | fs | 21.98 | 0.16 | 9.1 | 20.3 | 22.2 | s | 20.56 | 0.05 |
| \multicolumn{12}{c}{QSO 1416+067, $z = 1.436$, seeing=1.9 arcsec, magnitude limit=22.58} |
| -8.0 | -12.0 | 14.4 | g | 21.02 | 0.06 | 0.0 | 0.0 | 0.0 | Q | 16.44 | 0.00 |
| -6.3 | 5.2 | 8.2 | fg | 21.93 | 0.11 | 10.6 | -23.7 | 25.9 | fs | 22.52 | 0.18 |
| -4.4 | -13.9 | 14.5 | s | 16.64 | 0.00 | | | | | | |

**Table 2.** – continued

| ΔRA | ΔDec | sep | cls | $m_R$ | $\sigma_{m_R}$ | ΔRA | ΔDec | sep | cls | $m_R$ | $\sigma_{m_R}$ |
|---|---|---|---|---|---|---|---|---|---|---|---|
| | | | | | | | | | | | |
| | | QSO 1418+255, $z = 1.050$, seeing=2.6 arcsec, magnitude limit=21.90 | | | | | | | | | |
| -0.5 | -28.8 | 28.8 | g | 19.54 | 0.03 | 0.0 | 0.0 | 0.0 | Q | 16.13 | 0.00 |
| | | | | | | | | | | | |
| | | QSO 1421+122, $z = 1.611$, seeing=1.9 arcsec, magnitude limit=22.40 | | | | | | | | | |
| 0.1 | 0.0 | 0.1 | Q | 17.94 | 0.00 | 10.7 | -3.9 | 11.4 | s | 21.38 | 0.09 |
| 2.9 | -6.0 | 6.7 | s | 20.09 | 0.03 | 16.6 | -2.0 | 16.7 | g | 21.29 | 0.08 |
| 5.6 | 6.6 | 8.7 | fs | 22.04 | 0.15 | 20.1 | 2.0 | 20.2 | fs | 22.01 | 0.16 |
| 7.3 | -11.4 | 13.6 | g | 21.28 | 0.07 | | | | | | |
| | | | | | | | | | | | |
| | | QSO 1421+330, $z = 1.904$, seeing=1.8 arcsec, magnitude limit=22.02 | | | | | | | | | |
| -4.1 | -4.8 | 6.3 | g | 21.65 | 0.15 | 8.7 | -7.1 | 11.2 | fs | 22.61 | 0.32 |
| 0.0 | 0.0 | 0.0 | Q | 15.98 | 0.00 | | | | | | |
| | | | | | | | | | | | |
| | | QSO 1435+638, $z = 2.068$, seeing=2.0 arcsec, magnitude limit=21.59 | | | | | | | | | |
| 0.0 | 0.0 | 0.0 | Q | 16.11 | 0.00 | 18.2 | 21.5 | 28.2 | g | 20.17 | 0.06 |
| 1.1 | 22.5 | 22.5 | fs | 21.68 | 0.23 | | | | | | |
| | | | | | | | | | | | |
| | | QSO 1511+103, $z = 1.546$, seeing=2.7 arcsec, magnitude limit=21.04 | | | | | | | | | |
| -26.6 | 11.9 | 29.2 | fs | 21.35 | 0.25 | -4.5 | 20.4 | 20.9 | s | 19.38 | 0.04 |
| -19.0 | -7.7 | 20.5 | s | 17.13 | 0.01 | 0.0 | 0.0 | 0.0 | Q | 16.79 | 0.01 |
| -12.7 | -19.6 | 23.4 | g | 19.65 | 0.06 | 2.4 | -26.9 | 27.0 | fs | 21.04 | 0.26 |
| | | | | | | | | | | | |
| | | QSO 1512+370, $z = 0.371$, seeing=1.9 arcsec, magnitude limit=21.32 | | | | | | | | | |
| -25.5 | -11.6 | 28.0 | fs | 21.15 | 0.18 | -5.1 | 3.7 | 6.2 | fg | 21.32 | 0.25 |
| -24.2 | 1.9 | 24.3 | g | 20.75 | 0.12 | 0.1 | 0.0 | 0.1 | Q | 16.20 | 0.00 |
| -20.1 | -10.7 | 22.7 | g | 20.72 | 0.11 | 10.0 | -2.9 | 10.4 | g | 19.96 | 0.06 |
| | | | | | | | | | | | |
| | | QSO 1517+239, $z = 1.898$, seeing=2.7 arcsec, magnitude limit=21.99 | | | | | | | | | |
| 0.0 | 0.0 | 0.0 | Q | 17.48 | 0.00 | 23.6 | -6.4 | 24.4 | fs | 22.92 | 0.58 |
| | | | | | | | | | | | |
| | | QSO 1556+335, $z = 1.646$, seeing=1.9 arcsec, magnitude limit=22.10 | | | | | | | | | |
| -21.2 | -7.2 | 22.4 | g | 21.15 | 0.10 | 13.4 | -23.0 | 26.6 | fg | 22.14 | 0.25 |
| -14.7 | 5.6 | 15.7 | fs | 21.58 | 0.15 | 19.6 | -21.7 | 29.2 | g | 20.20 | 0.04 |
| 0.0 | 0.0 | 0.0 | Q | 17.20 | 0.00 | 20.4 | -8.6 | 22.2 | g | 20.60 | 0.06 |
| 0.6 | 14.5 | 14.5 | fs | 22.16 | 0.24 | | | | | | |
| | | | | | | | | | | | |
| | | QSO 1559+173, $z = 1.944$, seeing=2.1 arcsec, magnitude limit=22.30 | | | | | | | | | |
| -28.6 | 1.7 | 28.6 | s | 19.24 | 0.01 | 8.5 | 15.9 | 18.0 | fs | 21.99 | 0.14 |
| -18.9 | -13.6 | 23.3 | s | 13.30 | SAT | 10.5 | -11.7 | 15.7 | s | 18.43 | 0.01 |
| -14.6 | 1.7 | 14.7 | g | 21.13 | 0.10 | 11.2 | -17.1 | 20.4 | s | 21.40 | 0.09 |
| -10.8 | -5.3 | 12.1 | fg | 22.30 | 0.29 | 13.9 | -26.5 | 29.9 | fg | 22.03 | 0.15 |
| -10.6 | -3.1 | 11.1 | fg | 22.74 | 0.46 | 20.2 | -21.5 | 29.5 | fs | 22.15 | 0.17 |
| -2.0 | 27.5 | 27.6 | g | 20.94 | 0.06 | 25.6 | 11.5 | 28.1 | s | 17.09 | 0.00 |
| 0.0 | 0.0 | 0.0 | Q | 18.21 | 0.01 | | | | | | |
| | | | | | | | | | | | |
| | | QSO 1602−001, $z = 1.625$, seeing=2.2 arcsec, magnitude limit=22.32 | | | | | | | | | |
| -24.2 | 17.1 | 29.7 | g | 20.72 | 0.06 | 3.8 | -20.6 | 20.9 | s | 18.53 | 0.01 |
| -17.8 | 6.4 | 19.0 | s | 17.57 | 0.00 | 13.1 | -20.1 | 24.0 | s | 20.51 | 0.04 |
| -10.2 | 27.9 | 29.8 | fs | 23.12 | 0.43 | 14.2 | 21.2 | 25.5 | fs | 22.45 | 0.28 |
| -6.8 | 3.7 | 7.7 | s | 19.28 | 0.01 | 18.3 | -21.9 | 28.5 | fs | 22.36 | 0.22 |
| 0.0 | 0.0 | 0.0 | Q | 17.19 | 0.00 | 19.9 | 19.7 | 28.0 | s | 19.15 | 0.01 |
| | | | | | | | | | | | |
| | | QSO 1621+391, $z = 1.980$, seeing=2.0 arcsec, magnitude limit=22.30 | | | | | | | | | |
| 0.0 | 0.0 | 0.1 | Q | 19.53 | 0.02 | 14.0 | 25.7 | 29.2 | fs | 21.53 | 0.14 |
| 3.9 | -0.8 | 3.9 | fg | 21.41 | 0.13 | | | | | | |

**Table 2.** – continued

| ΔRA | ΔDec | sep | cls | $m_R$ | $\sigma_{m_R}$ | ΔRA | ΔDec | sep | cls | $m_R$ | $\sigma_{m_R}$ |
|---|---|---|---|---|---|---|---|---|---|---|---|
| \multicolumn{12}{c}{QSO 1621+412, $z = 1.620$, seeing=1.7 arcsec, magnitude limit=22.44} |
| -3.1 | -17.2 | 17.4 | s | 20.88 | 0.05 | 15.6 | 9.8 | 18.4 | g | 21.47 | 0.08 |
| 0.0 | 0.0 | 0.0 | Q | 16.71 | 0.00 | 16.6 | 7.3 | 18.1 | fg | 22.44 | 0.20 |
| 5.6 | 14.7 | 15.8 | g | 21.75 | 0.10 | 17.4 | -8.0 | 19.1 | g | 21.51 | 0.08 |
| \multicolumn{12}{c}{QSO 1630+374, $z = 1.470$, seeing=2.8 arcsec, magnitude limit=21.91} |
| -20.9 | -10.4 | 23.3 | s | 17.53 | 0.00 | 0.0 | 0.0 | 0.0 | Q | 15.85 | 0.00 |
| -8.7 | 4.3 | 9.7 | g | 20.94 | 0.08 | 14.6 | 2.7 | 14.8 | g | 19.57 | 0.02 |
| -0.6 | 14.9 | 14.9 | s | 16.92 | 0.00 | | | | | | |
| \multicolumn{12}{c}{QSO 1631+393, $z = 1.000$, seeing=2.1 arcsec, magnitude limit=22.36} |
| 0.0 | 0.0 | 0.0 | Q | 17.56 | 0.00 | 17.2 | -23.4 | 29.0 | s | 21.49 | 0.11 |
| \multicolumn{12}{c}{QSO 1634+176, $z = 1.897$, seeing=1.8 arcsec, magnitude limit=22.13} |
| -26.7 | 3.9 | 27.0 | fs | 22.21 | 0.25 | 2.2 | 0.4 | 2.3 | g | 21.44 | 0.11 |
| -23.5 | 0.7 | 23.5 | g | 21.02 | 0.08 | 3.6 | 18.7 | 19.0 | s | 21.41 | 0.11 |
| -19.2 | 18.7 | 26.8 | fs | 22.40 | 0.25 | 8.9 | 10.5 | 13.8 | fs | 21.95 | 0.19 |
| -11.0 | 27.9 | 30.0 | s | 18.86 | 0.01 | 11.5 | 0.9 | 11.5 | s | 14.78 | 0.00 |
| 0.0 | 0.0 | 0.0 | Q | 18.57 | 0.01 | 20.5 | 20.9 | 29.2 | s | 18.13 | 0.01 |
| 0.6 | 14.6 | 14.6 | fs | 21.66 | 0.14 | 25.1 | -3.9 | 25.4 | fg | 22.59 | 0.31 |
| \multicolumn{12}{c}{QSO 1634+706, $z = 1.334$, seeing=1.9 arcsec, magnitude limit=21.73} |
| -16.7 | 16.6 | 23.6 | fg | 20.90 | 0.10 | 0.4 | 28.5 | 28.5 | s | 17.33 | 0.01 |
| 0.0 | 0.0 | 0.0 | Q | 13.67 | 0.00 | 18.1 | 20.7 | 27.5 | fg | 22.14 | 0.31 |
| \multicolumn{12}{c}{QSO 1641+399, $z = 0.595$, seeing=1.8 arcsec, magnitude limit=22.09} |
| -20.7 | 10.6 | 23.3 | fs | 21.42 | 0.10 | -0.1 | 0.0 | 0.1 | Q | 17.44 | 0.00 |
| -16.3 | -15.4 | 22.5 | g | 21.33 | 0.10 | 6.6 | -1.2 | 6.8 | g | 21.59 | 0.15 |
| -11.0 | 27.8 | 29.9 | fg | 21.58 | 0.13 | 17.3 | 6.2 | 18.3 | s | 19.74 | 0.02 |
| \multicolumn{12}{c}{QSO 1704+608, $z = 0.371$, seeing=2.4 arcsec, magnitude limit=21.14} |
| -2.7 | 15.5 | 15.8 | s | 18.78 | 0.03 | 8.9 | -25.7 | 27.2 | fs | 20.37 | 0.10 |
| 0.0 | 0.0 | 0.0 | Q | 14.90 | 0.00 | 9.0 | -27.5 | 29.0 | fs | 20.44 | 0.11 |
| 3.0 | 6.6 | 7.3 | g | 20.45 | 0.12 | 25.5 | 15.6 | 29.9 | fs | 20.85 | 0.15 |
| \multicolumn{12}{c}{QSO 1715+535, $z = 1.929$, seeing=1.7 arcsec, magnitude limit=21.77} |
| -22.8 | 16.8 | 28.3 | g | 20.72 | 0.07 | 3.4 | 0.8 | 3.5 | s | 15.48 | 0.02 |
| -19.9 | -20.3 | 28.4 | g | 21.31 | 0.12 | 12.8 | 24.9 | 28.1 | s | 17.88 | 0.01 |
| 0.0 | 0.0 | 0.0 | Q | 15.40 | 0.02 | 19.7 | -1.4 | 19.8 | g | 19.74 | 0.03 |
| \multicolumn{12}{c}{QSO 1756+237, $z = 1.721$, seeing=2.5 arcsec, magnitude limit=21.13} |
| -20.8 | 11.5 | 23.8 | fg | 21.07 | 0.18 | 0.0 | 0.1 | 0.1 | Q | 16.21 | 0.00 |
| -19.2 | 11.4 | 22.3 | fs | 20.88 | 0.15 | 0.2 | -6.5 | 6.5 | fg | 21.32 | 0.26 |
| -9.4 | -28.2 | 29.7 | fs | 21.38 | 0.25 | 1.5 | 20.0 | 20.1 | fs | 21.02 | 0.19 |
| -6.4 | 26.8 | 27.6 | g | 20.71 | 0.14 | 2.3 | 12.9 | 13.0 | fs | 21.64 | 0.36 |
| -6.1 | 9.1 | 11.0 | s | 17.08 | 0.01 | 7.1 | -1.5 | 7.2 | s | 18.23 | 0.01 |
| -3.4 | -2.3 | 4.1 | g | 19.63 | 0.05 | 11.4 | 24.3 | 26.8 | s | 19.26 | 0.04 |
| -1.2 | 4.4 | 4.6 | s | 17.16 | 0.01 | 13.0 | 1.3 | 13.2 | g | 19.48 | 0.04 |
| -0.3 | 22.9 | 22.9 | s | 20.76 | 0.14 | 22.2 | -10.4 | 24.5 | s | 18.08 | 0.01 |
| \multicolumn{12}{c}{QSO 1828+487, $z = 0.692$, seeing=2.9 arcsec, magnitude limit=21.74} |
| -23.3 | -17.6 | 29.2 | g | 20.74 | 0.07 | 19.7 | -2.3 | 19.9 | fg | 21.67 | 0.16 |
| -5.6 | -7.3 | 9.2 | s | 20.08 | 0.04 | 26.3 | -1.9 | 26.4 | s | 16.56 | 0.00 |
| 0.0 | 0.0 | 0.0 | Q | 16.59 | 0.00 | | | | | | |

**Table 2.** – continued

| ΔRA | ΔDec | sep | cls | $m_R$ | $\sigma_{m_R}$ | ΔRA | ΔDec | sep | cls | $m_R$ | $\sigma_{m_R}$ |
|---|---|---|---|---|---|---|---|---|---|---|---|
| | | | | | QSO 1830+285, $z = 0.594$, seeing=1.6 arcsec, magnitude limit=21.74 | | | | | | |
| -15.3 | 22.8 | 27.4 | fs | 21.74 | 0.20 | 3.2 | -10.8 | 11.2 | s | 19.44 | 0.03 |
| -11.3 | -13.2 | 17.4 | s | 21.14 | 0.11 | 17.0 | -14.8 | 22.6 | s | 17.42 | 0.10 |
| -10.2 | -4.2 | 11.1 | s | 20.60 | 0.07 | 17.2 | -8.0 | 18.9 | s | 17.47 | 0.01 |
| -0.5 | 20.3 | 20.3 | s | 18.50 | 0.01 | 17.6 | -16.6 | 24.2 | s | 16.61 | 0.03 |
| -0.1 | 0.0 | 0.1 | Q | 18.12 | 0.01 | | | | | | |
| | | | | | QSO 1901+319, $z = 0.635$, seeing=2.3 arcsec, magnitude limit=21.34 | | | | | | |
| -24.6 | 16.7 | 29.7 | s | 20.99 | 0.17 | -2.2 | -2.2 | 3.1 | s | 17.15 | 0.01 |
| -21.5 | 9.5 | 23.5 | g | 20.15 | 0.08 | 0.0 | 0.0 | 0.0 | Q | 16.34 | 0.01 |
| -20.2 | 21.1 | 29.2 | s | 20.17 | 0.08 | 3.7 | 8.9 | 9.6 | fs | 21.61 | 0.30 |
| -17.5 | -0.6 | 17.5 | s | 15.05 | 0.00 | 3.9 | 27.6 | 27.8 | s | 20.51 | 0.11 |
| -16.3 | 15.9 | 22.8 | g | 20.57 | 0.12 | 5.5 | -1.3 | 5.7 | g | 20.30 | 0.09 |
| -14.5 | -14.0 | 20.2 | g | 19.70 | 0.06 | 6.6 | 18.2 | 19.4 | s | 21.26 | 0.20 |
| -14.3 | 6.7 | 15.8 | g | 19.53 | 0.04 | 8.1 | -27.3 | 28.5 | s | 16.21 | 0.00 |
| -13.6 | 22.9 | 26.6 | g | 20.62 | 0.12 | 10.0 | 8.5 | 13.1 | g | 20.48 | 0.10 |
| -9.0 | -13.5 | 16.3 | s | 14.91 | 0.00 | 10.5 | 1.0 | 10.5 | s | 18.51 | 0.02 |
| -8.0 | 21.9 | 23.3 | s | 18.16 | 0.01 | 15.5 | -16.2 | 22.5 | s | 18.63 | 0.02 |
| -6.4 | 25.1 | 25.9 | s | 19.50 | 0.04 | 17.1 | -3.7 | 17.5 | s | 18.57 | 0.02 |
| -5.8 | -7.4 | 9.4 | g | 20.00 | 0.08 | 17.2 | -22.2 | 28.1 | s | 19.67 | 0.05 |
| -4.5 | -25.2 | 25.6 | s | 18.01 | 0.01 | 23.0 | 10.8 | 25.4 | s | 17.85 | 0.01 |
| -4.4 | 18.4 | 18.9 | s | 18.40 | 0.01 | | | | | | |
| | | | | | QSO 2120+168, $z = 1.805$, seeing=2.2 arcsec, magnitude limit=21.53 | | | | | | |
| -24.4 | 4.6 | 24.9 | s | 18.90 | 0.02 | 1.2 | -7.0 | 7.1 | fg | 21.87 | 0.27 |
| -14.2 | -4.7 | 14.9 | s | 19.43 | 0.03 | 8.2 | 0.3 | 8.2 | fs | 21.46 | 0.17 |
| -10.2 | -1.0 | 10.2 | s | 18.03 | 0.01 | 21.5 | 17.1 | 27.4 | fs | 22.23 | 0.37 |
| -5.0 | -15.3 | 16.1 | s | 18.92 | 0.02 | 21.7 | 3.6 | 22.0 | fs | 22.18 | 0.32 |
| -0.1 | 0.0 | 0.1 | Q | 17.38 | 0.00 | | | | | | |
| | | | | | QSO 2145+067, $z = 0.990$, seeing=2.8 arcsec, magnitude limit=20.94 | | | | | | |
| -11.9 | 24.4 | 27.2 | fs | 23.06 | 1.51 | 0.0 | 0.0 | 0.0 | Q | 15.39 | 0.00 |
| -2.0 | 25.7 | 25.8 | fs | 20.98 | 0.21 | | | | | | |
| | | | | | QSO 2156+297, $z = 1.753$, seeing=2.5 arcsec, magnitude limit=21.60 | | | | | | |
| -22.2 | 14.3 | 26.4 | fs | 21.36 | 0.17 | 6.2 | 6.3 | 8.9 | fg | 20.94 | 0.12 |
| -13.4 | 25.0 | 28.4 | s | 19.20 | 0.02 | 10.2 | 21.3 | 23.6 | g | 19.78 | 0.04 |
| -10.0 | 27.7 | 29.5 | fg | 20.63 | 0.08 | 20.1 | 13.4 | 24.1 | s | 16.68 | 0.00 |
| -0.1 | 0.0 | 0.1 | Q | 19.36 | 0.03 | 22.8 | -9.8 | 24.8 | s | 15.07 | 0.00 |
| | | | | | QSO 2230+114, $z = 1.037$, seeing=2.1 arcsec, magnitude limit=22.58 | | | | | | |
| -26.4 | 14.2 | 30.0 | g | 21.62 | 0.07 | 2.6 | -14.3 | 14.6 | fg | 23.20 | 0.33 |
| -13.0 | 5.7 | 14.2 | g | 22.26 | 0.14 | 13.2 | 25.1 | 28.3 | s | 19.79 | 0.01 |
| 0.0 | 0.1 | 0.1 | Q | 16.63 | 0.00 | 29.0 | 5.4 | 29.5 | s | 22.21 | 0.14 |

### Notes to Table 3

The columns are as follows: 1,2: position relative to quasar in arcseconds of RA and Dec, respectively, 3: separation from quasar in arcseconds, 4: classification—galaxy (g), faint galaxy (fg), star (s), faint star (fs) or quasar (Q). Categories fg and fs are objects selected at a fixed instrumental magnitude whose classification we consider to be very uncertain. 5: R magnitude, 6: relative measure of S/N of detection (see Paper 1), 7-12: as 1-6 for additional objects. SAT means the stellar object was saturated.

**Table 3.** Nearest-neighbour galaxy data for each quasar field.

| (1) QSO | (2) $z_q$ | (3) mlim | (4) mqso | (5) mgal | (6) area | (7) $N$ | (8) sep | (9) $s$ | (10) $L/L_*$ |
|---|---|---|---|---|---|---|---|---|---|
| 0710+118 | 0.768 | 21.22 | 16.05 | 19.23 | 3.32 | 12 | 28.74 | 0.623 | 13.53 |
| 0736−063 | 1.914 | 21.52 | 16.73 | 20.10 | 3.37 | 21 | 3.88 | 0.029 | 99.99 |
| 0742+318 | 0.462 | 22.50 | 15.65 | 22.36 | 3.34 | 23 | 19.73 | 0.576 | 0.17 |
| 0835+580 | 1.534 | 21.89 | 16.85 | 21.58 | 3.34 | 16 | 6.87 | 0.069 | 18.62 |
| 0836+195 | 1.691 | 22.36 | 17.23 | 22.06 | 3.33 | 10 | 10.46 | 0.099 | 17.61 |
| 0838+355 | 2.530 | 21.78 | 16.46 | 21.61 | 3.34 | 17 | 20.59 | 0.500 | 99.99 |
| 0843+136 | 1.875 | 21.88 | 17.10 | 20.17 | 3.34 | 17 | 16.12 | 0.343 | 99.99 |
| 0848+163 | 1.925 | 22.18 | 17.36 | 19.21 | 3.33 | 15 | 15.19 | 0.281 | 99.99 |
| 0854+191 | 1.896 | 22.21 | 17.60 | 21.11 | 3.31 | 11 | 35.13 | 0.746 | 67.99 |
| 0856+170 | 1.449 | 22.15 | 17.78 | 21.45 | 3.34 | 14 | 10.40 | 0.133 | 17.16 |
| 0856+189 | 1.286 | 22.00 | 17.67 | 20.55 | 3.35 | 13 | 45.21 | 0.937 | 26.19 |
| 0935+414 | 1.940 | 21.82 | 15.89 | 19.42 | 3.34 | 12 | 25.26 | 0.524 | 99.99 |
| 0952+179 | 1.472 | 21.93 | 17.20 | 21.00 | 3.38 | 9 | 22.59 | 0.354 | 27.36 |
| 0955+326 | 0.533 | 22.06 | 15.96 | 21.85 | 3.35 | 5 | 26.55 | 0.289 | 0.39 |
| 0958+551 | 1.751 | 22.47 | 15.97 | 19.30 | 3.12 | 16 | 23.42 | 0.597 | 99.99 |
| 1017+280 | 1.928 | 22.16 | 15.73 | 20.59 | 3.37 | 12 | 18.14 | 0.312 | 99.99 |
| 1038+064 | 1.270 | 21.53 | 16.26 | 21.03 | 3.34 | 11 | 9.72 | 0.093 | 16.10 |
| 1054−034 | 2.115 | 22.27 | 17.82 | 19.93 | 3.34 | 18 | 5.67 | 0.053 | 99.99 |
| 1055−045 | 1.428 | 22.46 | 17.27 | 22.17 | 3.33 | 18 | 35.08 | 0.891 | 8.42 |
| 1103−006 | 0.426 | 21.93 | 16.35 | 19.71 | 3.32 | 18 | 14.88 | 0.317 | 1.57 |
| 1104+167 | 0.634 | 21.83 | 16.08 | 21.51 | 3.33 | 16 | 14.12 | 0.262 | 0.89 |
| 1115+080 | 1.718 | 22.40 | 15.89 | 20.33 | 3.35 | 21 | 12.33 | 0.260 | 92.40 |
| 1126+101 | 1.515 | 22.02 | 17.07 | 20.82 | 3.34 | 17 | 7.67 | 0.090 | 35.73 |
| 1138+040 | 1.877 | 22.20 | 16.86 | 21.08 | 3.36 | 27 | 17.16 | 0.530 | 66.96 |
| 1146+372 | 0.800 | 22.54 | 17.55 | 21.67 | 3.34 | 16 | 10.78 | 0.161 | 1.65 |
| 1148−001 | 1.982 | 22.12 | 16.99 | 21.01 | 3.35 | 9 | 29.45 | 0.534 | 90.31 |
| 1209+107 | 2.191 | 22.22 | 17.63 | 21.43 | 3.34 | 21 | 7.19 | 0.097 | 96.11 |
| 1211+334 | 1.598 | 22.32 | 17.32 | 21.47 | 3.34 | 20 | 22.00 | 0.606 | 24.18 |
| 1215+113 | 1.396 | 22.27 | 16.69 | 20.96 | 3.33 | 8 | 46.54 | 0.839 | 23.80 |
| 1225+317 | 2.219 | 22.49 | 15.52 | 21.89 | 3.36 | 10 | 38.79 | 0.780 | 66.71 |
| 1227+292 | 0.740 | 22.37 | 16.87 | 21.93 | 3.34 | 14 | 5.98 | 0.046 | 0.99 |
| 1228+078 | 1.813 | 22.12 | 17.33 | 20.99 | 3.37 | 14 | 16.87 | 0.314 | 62.84 |
| 1231+292 | 2.000 | 22.26 | 16.66 | 19.13 | 3.34 | 11 | 29.91 | 0.620 | 99.99 |
| 1237+280 | 1.830 | 22.55 | 18.15 | 21.82 | 3.28 | 12 | 37.59 | 0.826 | 30.43 |
| 1246−057 | 2.236 | 22.01 | 16.08 | 20.60 | 3.35 | 18 | 41.52 | 0.958 | 99.99 |
| 1246+377 | 1.242 | 22.31 | 16.89 | 21.04 | 3.33 | 10 | 66.49 | 0.995 | 14.75 |
| 1256+357 | 1.882 | 22.58 | 17.60 | 22.06 | 3.30 | 28 | 7.40 | 0.136 | 27.46 |
| 1257+346 | 1.375 | 22.02 | 16.38 | 21.32 | 3.27 | 15 | 17.39 | 0.357 | 16.22 |
| 1258+283 | 1.360 | 22.24 | 16.77 | 21.61 | 3.34 | 21 | 26.60 | 0.765 | 11.97 |
| 1258+286 | 1.922 | 21.98 | 17.29 | 21.09 | 3.37 | 13 | 24.82 | 0.536 | 73.42 |

**Table 3.** – continued

| (1) | (2) | (3) | (4) | (5) | (6) | (7) | (8) | (9) | (10) |
|-----|-----|-----|-----|-----|-----|-----|-----|-----|------|
| QSO | $z_q$ | mlim | mqso | mgal | area | $N$ | sep | $s$ | $L/L_*$ |
| 1303+308 | 1.759 | 22.33 | 17.24 | 20.63 | 3.33 | 17 | 19.90 | 0.477 | 77.21 |
| 1308+182 | 1.677 | 22.62 | 18.55 | 21.06 | 3.33 | 25 | 29.49 | 0.882 | 42.76 |
| 1309−056 | 2.224 | 21.95 | 17.28 | 21.33 | 3.34 | 23 | 16.17 | 0.436 | 99.99 |
| 1309+340 | 1.035 | 22.61 | 17.33 | 21.27 | 3.07 | 17 | 7.03 | 0.083 | 6.14 |
| 1317+277 | 1.022 | 21.95 | 16.13 | 20.41 | 3.33 | 8 | 34.90 | 0.623 | 12.93 |
| 1318+29a | 1.703 | 22.01 | 17.13 | 21.02 | 3.35 | 5 | 47.96 | 0.703 | 47.22 |
| 1318+29b | 0.549 | 22.01 | 16.30 | 21.55 | 3.35 | 5 | 46.60 | 0.679 | 0.56 |
| 1329+412 | 1.935 | 21.91 | 17.03 | 20.24 | 3.34 | 12 | 30.14 | 0.657 | 99.99 |
| 1331+170 | 2.081 | 22.23 | 16.46 | 20.96 | 3.35 | 12 | 27.50 | 0.586 | 99.99 |
| 1346−036 | 2.344 | 22.58 | 16.86 | 21.73 | 3.36 | 21 | 25.05 | 0.719 | 99.97 |
| 1354+195 | 0.720 | 22.40 | 16.31 | 21.73 | 3.33 | 14 | 7.12 | 0.065 | 1.09 |
| 1416+067 | 1.436 | 22.58 | 16.44 | 21.93 | 3.36 | 35 | 8.16 | 0.196 | 10.70 |
| 1416+159 | 1.472 | 22.23 | 17.77 | 20.95 | 3.35 | 9 | 22.49 | 0.355 | 28.65 |
| 1418+255 | 1.050 | 21.90 | 16.13 | 19.54 | 3.34 | 8 | 28.83 | 0.479 | 31.87 |
| 1421+122 | 1.611 | 22.40 | 17.94 | 21.28 | 3.36 | 23 | 13.56 | 0.329 | 29.74 |
| 1421+330 | 1.904 | 22.02 | 15.98 | 21.65 | 3.33 | 8 | 6.29 | 0.030 | 42.10 |
| 1435+638 | 2.068 | 21.59 | 16.11 | 20.17 | 3.28 | 9 | 28.18 | 0.510 | 99.99 |
| 1511+103 | 1.546 | 21.04 | 16.79 | 19.65 | 3.31 | 6 | 23.40 | 0.274 | 99.99 |
| 1512+370 | 0.371 | 21.32 | 16.20 | 19.96 | 3.32 | 14 | 10.44 | 0.135 | 0.89 |
| 1517+239 | 1.898 | 21.99 | 17.48 | 19.95 | 3.35 | 10 | 35.33 | 0.712 | 99.99 |
| 1556+335 | 1.646 | 22.10 | 17.20 | 20.60 | 3.28 | 10 | 22.16 | 0.383 | 60.61 |
| 1559+173 | 1.944 | 22.30 | 18.21 | 21.13 | 3.11 | 18 | 14.66 | 0.327 | 74.33 |
| 1602−001 | 1.625 | 22.32 | 17.19 | 20.72 | 3.21 | 15 | 29.69 | 0.742 | 51.56 |
| 1621+391 | 1.980 | 22.30 | 19.53 | 21.41 | 3.34 | 3 | 3.94 | 0.004 | 62.20 |
| 1621+412 | 1.620 | 22.44 | 16.71 | 21.75 | 3.34 | 25 | 15.76 | 0.447 | 19.72 |
| 1630+374 | 1.470 | 21.91 | 15.85 | 20.94 | 3.31 | 22 | 9.71 | 0.179 | 28.79 |
| 1631+393 | 1.000 | 22.36 | 17.56 | 22.08 | 3.34 | 5 | 39.00 | 0.538 | 2.56 |
| 1634+176 | 1.897 | 22.13 | 18.57 | 21.44 | 3.32 | 30 | 2.27 | 0.015 | 50.29 |
| 1634+706 | 1.334 | 21.73 | 13.67 | 20.90 | 3.34 | 19 | 23.58 | 0.639 | 21.54 |
| 1641+399 | 0.595 | 22.09 | 17.44 | 21.59 | 3.34 | 23 | 6.75 | 0.094 | 0.68 |
| 1704+608 | 0.371 | 21.14 | 14.90 | 20.45 | 3.34 | 11 | 7.27 | 0.053 | 0.57 |
| 1715+535 | 1.929 | 21.77 | 15.40 | 19.74 | 3.34 | 30 | 19.76 | 0.674 | 99.99 |
| 1756+237 | 1.721 | 21.13 | 16.21 | 19.63 | 3.08 | 26 | 4.05 | 0.043 | 99.99 |
| 1828+487 | 0.692 | 21.74 | 16.59 | 21.67 | 3.35 | 18 | 19.89 | 0.494 | 1.01 |
| 1830+285 | 0.594 | 21.74 | 18.12 | 20.38 | 3.34 | 15 | 47.43 | 0.972 | 2.06 |
| 1901+319 | 0.635 | 21.34 | 16.34 | 20.30 | 3.34 | 31 | 5.66 | 0.089 | 2.72 |
| 2120+168 | 1.805 | 21.69 | 17.38 | 20.64 | 3.34 | 13 | 37.77 | 0.846 | 85.16 |
| 2145+067 | 0.990 | 20.94 | 15.39 | 19.14 | 3.31 | 14 | 50.51 | 0.979 | 36.98 |
| 2156+297 | 1.753 | 21.60 | 19.36 | 20.94 | 3.34 | 23 | 8.89 | 0.158 | 57.22 |
| 2230+114 | 1.037 | 22.58 | 16.63 | 22.26 | 3.37 | 23 | 14.21 | 0.354 | 2.48 |

Notes to Table 3

1: quasar name, 2: quasar redshift, 3: limiting magnitude, 4: measured quasar R magnitude, 5: measured R magnitude of nng, 6: useful area of field ($/10^4$ arcsec$^2$), 7: number of galaxies on field, 8: separation of quasar and nng ($/$arcsec), 9: $s$, 10: luminosity of nng if at quasar redshift.

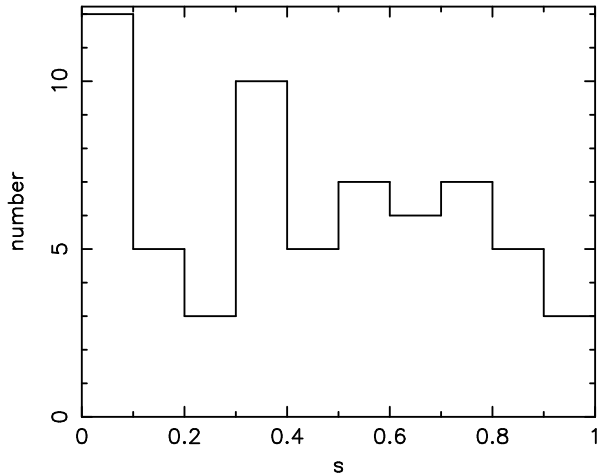

**Figure 1.** Histograms of the values of the s-parameter for the sample, $z_q > 1.0$

test which looks at the maximum deviation of the cumulative distribution of s-values as compared with a linear distribution. This has the advantage that it is non-parametric but unfortunately it is not particularly sensitive to the kinds of fluctuations we expect. It is more useful as a general check that our analysis of the star distribution and of objects around control stars does not show any peculiarities which might indicate an error in our procedures. In no case does it give a significant result and we do not report the results here. To test specifically for nearby neighbours we split the sample into two bins with $s \leq 0.1$ and $s > 0.1$. Then we calculate the probability of getting the observed number (or more) objects in the smaller bin, given as our null hypothesis that the distribution is Binomial with probabilities 0.1 and 0.9 of objects appearing in each bin, respectively.

The results are shown in Table 4 for both galaxies and stars and for various magnitude cuts, as indicated in the first two columns. Column 2 shows which of the various criteria has been used to try and eliminate fields in which the nng may be at the quasar redshift. The simplest of these is to limit the sample to quasars with redshifts greater than unity; others will be discussed below. We also reject fields in which there are fewer than 3 detected objects (in addition to the quasar) although this only occurs at bright magnitude limits: the third column gives the number of fields which survive these constraints. Next come the number of fields for which $s < 0.1$ and the Binomial probability that this is consistent with a uniform distribution. Columns 6–8 will be described in Section 3.3, below.

Concentrating on the full sample we find that the hypothesis that the galaxy distribution is random is rejected at the 2.1 percent confidence level. Figure 1 shows that there is a significant concentration of galaxies at low values of $s$, i.e. small separations. This is an extremely robust conclusion and we can see no way of escaping from it.

We next discuss several aspects of the quasar-galaxy distribution in more detail:

(i) One obvious explanation for the observed association would be if the galaxies were clustered around the quasar itself. We have tried to avoid this possibility by rejecting all quasar fields where the quasar redshift is less than unity. However there is one field with a low value of $s$, 1309+340, in which the quasar redshift is quite low ($z_q = 1.035$) and the nng reasonably faint ($m_R = 21.27$). Can we really state with confidence that the galaxy is not physically associated with the quasar? One objective measure is the luminosity the nng would have if it were at the quasar redshift: we call this $L_q$. Its calculation is outlined in the Appendix of Paper 1. In the above case $L_q \approx 6L_*$. While this is large it is perhaps not impossibly so for the dominant galaxy in a cluster, say. We have tried imposing the more stringent conditions $L_q > 10L_*$ and $L_q > 30L_*$ with the results shown in Table 4. The effect of the first constraint is to admit two previously rejected fields while eliminating three others including the one mentioned above. The number of fields with $s < 0.1$ drops by one and the Binomial probability increases to 0.042. With the higher luminosity cut the probability rises still further to 0.054 and the associations become statistically insignificant. Note, however, that the number of close associations per field is exactly the same for the original $z_q > 1$ constraint: it is just that there are fewer admissible fields and so the significance has decreased. Finally, we ought to note that the procedure described here introduces a slight bias although once again it acts so as to strengthen our conclusions: we should really eliminate *all* galaxies fainter than the threshold (not just throw out those fields in which the nng galaxy is this faint) because otherwise they will add to the background counts and increase $s$.

(ii) We considered the possibility that the catalogues from which we drew our quasar sample were biased towards the inclusion of quasars which might have a nearby galaxy. For example the quasar 0957+561 had to be eliminated from our sample, as discussed in Section 2.1. In general, however, we would expect the observational bias to be towards isolated quasars because some selection criteria favour stellar images. To check for a bias we repeated our analysis on two subset of quasars: those in which there are no known MgII absorption systems (including those quasars for which there is no absorption line data), and those in which there has been an unsuccessful search for MgII absorption—this should bias against the presence of a nearby galaxy. The results shown in Table 4 do indeed show reduced significance but once again this is largely due to the reduced size of the sample. The detection of 7/39 nng with $s < 0.1$ in the latter case corresponds to 11.3/63, only slightly fewer than in the whole sample. We conclude that there is no evidence for a selection bias.

(iii) The significance of the associations varies as the magnitude limit is decreased. The faintest magnitude limit, imposed by the condition that the magnitude error be less than 0.2, is 22.58. As this is lowered so faint galaxies disappear from the catalogues and the values of $s$ de-

**Table 4.** Nearest-neighbour statistics.

| (1) sample | (2) constraints | (3) $N$ | (4) $N_{<0.1}$ | (5) prob | (6) $N_g$ | (7) range | (8) flp | (9) notes |
|---|---|---|---|---|---|---|---|---|
| $q-g$ | $z_q > 1.0$ | 63 | 12 | 0.021 | 6.3 | 1–11 | 0.46 | |
| $q-g$ | $L(z_q)/L_* > 10.$ | 62 | 11 | 0.042 | 5.3 | 0–10 | 0.45 | |
| $q-g$ | $L(z_q)/L_* > 30.$ | 42 | 8 | 0.054 | 4.2 | 0–8 | 0.41 | |
| $q-g$ | $z_q > 1.0$ | 39 | 7 | 0.089 | 3.4 | 0–7 | 0.49 | a |
| $q-g$ | $z_q > 1.0$ | 49 | 8 | 0.112 | 3.4 | 0–7 | 0.48 | b |
| $q-g$ | $z_q > 1.0$ | 32 | 8 | 0.012 | 5.3 | 1–8 | 0.46 | c |
| $q-g$ | $z_q > 1.0, m_R < \infty$ | 63 | 12 | 0.021 | 6.3 | 1–11 | 0.46 | |
| $q-g$ | $z_q > 1.0, m_R < 22.1$ | 63 | 13 | 0.009 | 7.4 | 2–12 | 0.44 | |
| $q-g$ | $z_q > 1.0, m_R < 22.0$ | 63 | 11 | 0.047 | 5.2 | 0–10 | 0.43 | |
| $q-g$ | $z_q > 1.0, m_R < 21.5$ | 62 | 9 | 0.163 | 3.1 | 0–8 | 0.36 | |
| $q-g$ | $z_q > 1.0, m_R < 21.0$ | 57 | 4 | 0.314 | −1.9 | 0–4 | 0.28 | |
| $q-g$ | $z_q > 1.0$ | 63 | 12 | 0.021 | 6.3 | 1–11 | 0.46 | |
| $c-g$ | $z_q > 1.0$ | 45 | 3 | 0.328 | −1.7 | 0–3 | 0.47 | |
| $q-s$ | $z_q > 1.0$ | 63 | 4 | 0.232 | −2.6 | 0–4 | 0.46 | |
| $c-s$ | $z_q > 1.0$ | 45 | 5 | 0.473 | 0.6 | 0–5 | 0.47 | |
| $q-g$ | $z_q > 1.0$ | 63 | 12 | 0.021 | 6.3 | 1–11 | 0.46 | |
| $q-g$ | $z_q > 1.5$ | 46 | 10 | 0.014 | 6.0 | 1–9 | 0.41 | |
| $q-g$ | $z_q > 1.0, M_q < -26.5$ | 48 | 9 | 0.046 | 4.7 | 0–8 | 0.44 | |
| $q-g$ | $z_q > 1.0$ | 63 | 12 | 0.021 | 6.3 | 1–11 | 0.46 | |
| $q-g$ | $z_q > 1.0$ | 63 | 12 | 0.021 | 6.3 | 2–11 | 0.46 | d |
| nng$-g$ | $z_q > 1.0$ | 62 | 6 | 0.573 | −0.2 | 0–5 | 0.46 | |
| nng$-s$ | $z_q > 1.0$ | 63 | 4 | 0.232 | −2.6 | 0–2 | 0.46 | |

## Notes to Table 4

1: sample—galaxies (g) or stars (s) around quasars (q), control stars(c) or nearest-neighbour galaxies (nng), 2: constraints imposed on the quasar redshift or absolute magnitude, on the redshift the nng would have if at the quasar redshift, or on the limiting magnitude, 3: number of useful fields, 4: number of nng (or nns) with $s < 0.1$, 5: Binomial probability of $N_{<0.1}$ given $N$, 6: number of excess galaxies estimated from $N$ and $N_{<0.1}$, 7: 95 percentile range for $N_g$, 8: fractional lensing path over which a galaxy with luminosity $0.4L_*$ could have been detected, 9: further notes as listed below, a: unsuccessful search for MgII in quasar spectra, b: unsuccessful or no search for MgII in quasar spectra, c: radio-loud quasars, d: area correction for hole under quasar image.

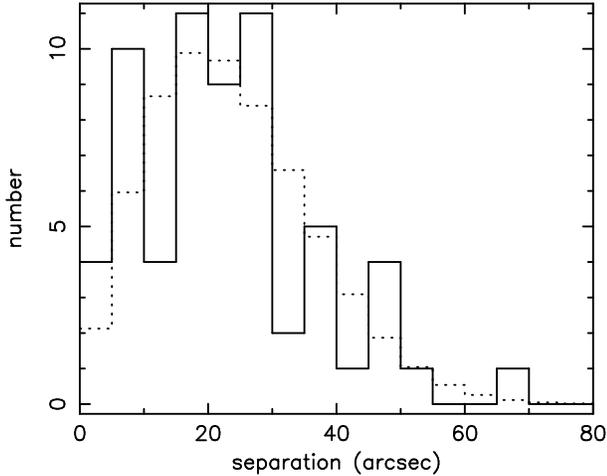

**Figure 2.** A histogram of the separations of the quasars and the nngs in physical units for the sample, $z_q > 1.0$

crease slightly (see Table 4). Eventually, however, the nng themselves are eliminated and the number with $s < 0.1$ drops. The significance of the associations peaks at $m_R = 22.1$ then declines rapidly at lower magnitude cuts. This suggests that many of the nng are at the faint end of the magnitude distribution (see Section 4.2). It also suggests that we may be missing many nng which are below the magnitude limits of our fields (some of which do not extend below $m_R = 21$): an attempt is made to correct for this in Section 3.3, below.

(iv) The distribution of quasar-nng separations for a magnitude limit $m_R = 22$ is shown in Figure 2. Also plotted is the Poisson distribution expected from the mean galaxy surface density averaged over all fields. These two distributions agree in the KS-test with a probability of 0.55. The excess of nng with separation less than 10 arcsec is significant at the 2.6 percent level, based on the assumption that the probability distribution is Binomial. However raising the background normalisation by just 25 percent increases these probabilities to 0.88 and 0.11, respectively. This emphasises the difficulty of using an external normalisation of the galaxy surface-density to try to detect excess galaxies.

(v) As a check on our results we also analysed the distribution of galaxies around control stars (as defined in Section 2.1). We show one set of results in Table 4 but this is representative of every case we have tried. There are fewer fields than for the quasars so we would not expect a significant result, however the number of nng with $s < 0.1$ is now less than one would expect for a random distribution. This is an indication that our sample is incomplete and is actually biased against the detection of galaxies near quasars. We will attempt to estimate the number of missing nng below but from this test alone it appears to be about one per thirty fields (with a large uncertainty).

(vi) A further check is provided by the distribution of stars around both quasars and control stars. These are also consistent with being random but once again show a small decrement in the number of nearest-neighbour stars (nns) with $s < 0.1$.

(vii) It is vital to our conclusions that the sensitivity to detection of faint objects is independent of position on the field. For example vignetting might lower the background galaxy count relative to that under the quasar and so give artificially low values of $s$. To some extent we have checked this by the adoption of control stars. However a further check is provided by the measurement of the surface density of galaxies and stars in different parts of the field, as shown in Table 5.

The mean number density of galaxies, avoiding the region within 30 arcsec of each quasar, is approximately $4.3 \times 10^{-4}$ galaxies per square arcsecond, both for the whole field and for the central region only (approximately one third of the total area). If we include the region around the quasar the equivalent numbers are still consistent with the original figure but go up slightly, reflecting the excess of nng at close separations.

The data for the stars is not so clear-cut. There does appear to be a marginally significantly greater density of stars near the centre of the frame rather than around the edge. This excess is limited to the bright stars, however, which have very large images whose radii are greater than the 10 pixel margin which we exclude from the edge of each field when compiling the image catalogues (see Section 2.2). The catalogues are therefore incomplete around the edge of the field for these bright objects. When we restrict the analysis to stars with $m_R \geq 18$ then they do appear to be uniformly distributed.

(viii) It is obvious that we cannot detect objects very close to the quasars and that this will adversely affect the statistics. One estimate of the missing numbers, taken from the number of detections around control stars given above, was 1 nng with $s < 0.1$ per 30 fields. This is consistent with number density data: if we look at the number density of nng around control stars then this shows a decrement at small separations (see Table 5). In 57 fields the number of missing galaxies is 3.7, 2.5, 1.6 and 1.0 within circles of radius 30, 20, 10 and 5 arcsec around the control star, respectively. None of these deficiencies is significant but they also suggest that we are missing one or two foreground galaxies from our sample of quasar fields. Adding two extra objects to the $s < 0.1$ bin would increase the significance of the association to $3.3 \times 10^{-3}$.

One way to correct for the missing area around quasars is to allow for it in the definition of $s$, as described in Section 3.1. The difficulty with this is in deciding how much area is hidden underneath the quasar. One crude estimate for galaxies is that the quasar-galaxy separation must be $r \gtrsim 16 - 0.75 m_R$ arcsec where $m_R$ is the measured quasar magnitude. This relation is taken from the locus of points in the $r$-$m_R$ plane but takes no account of seeing or limiting magnitude. Using this relation to correct the area no further nng are added to the $s < 0.1$ bin and so the significance does not change.

**Table 5.** Number density of objects to a magnitude limit of $m_R = 22$ on various parts of the fields, as described in the text.

| sample | region | number | density |
|---|---|---|---|
| galaxies | all | 1158 | 4.36 |
| 80 fields | central | 379 | 4.57 |
|  | all−qso | 1037 | 4.26 |
|  | central−qso | 270 | 4.28 |
| stars | all | 1946 | 7.39 |
| 80 fields | central | 639 | 7.70 |
|  | all−qso | 1778 | 7.31 |
|  | central−qso | 469 | 7.44 |
| stars | all | 1574 | 5.92 |
| 80 fields | central | 488 | 5.88 |
| $m_R \geq 18$ | all−qso | 1430 | 5.87 |
|  | central−qso | 360 | 5.71 |
| galaxies | < 30 arcsec | 65 | 4.03 |
| near control | < 20 arcsec | 28 | 3.91 |
| stars | < 10 arcsec | 6 | 3.35 |
| 57 fields | < 5 arcsec | 1 | 2.23 |
| galaxies | < 30 arcsec | 89 | 5.00 |
| near quasars | < 10 arcsec | 15 | 7.58 |
| with $z_q > 1.0$ | < 5 arcsec | 4 | 8.08 |
| 63 fields | < 3.53 arcsec | 1 | 4.05 |

Density units are $10^{-4}/\text{arcsec}^2$

(ix) Finally, we note that our original intention was to select a quasar sample which would be optimal for the detection of gravitational lensing, if it were present. The need for a large sample weakened this aim somewhat. We also analysed the data for two subsets closer to the original ideal and the results of these are also shown in Table 4. Restricting our attention to quasars with redshifts greater than 1.5 we find a more significant result (and a greater number of excess galaxies per quasar) whereas the more promising cut on absolute quasar magnitude, $M_R < -26.5$, gives a looser constraint (but about the same number of excess galaxies per quasar). We note that there is at least one field, 1115+080, where the quasar is multiply-imaged and the lensing galaxy is known, but is not detected by our analysis.

In conclusion we find a significant association of quasars and nearest-neighbour galaxies which depends solely on counting objects on our frames. Variations in the galaxy surface-density, seeing, magnitude limit, etc. make no difference to our conclusions. We have checked for a large variety of possible adverse biases, none of which appear to be present in our data. The inability to detect galaxies very close to the quasars leads to an underestimate of the number of close associations and strengthens our conclusions.

### 3.3 The number of excess galaxies

Having established that there is a population of galaxies associated with quasars, we next estimate how many quasars have such an observed excess galaxy. This number varies with the manner in which the sample is chosen but is about one tenth for the constraint, $z_q > 1.0$. After correcting for incompleteness around the quasar this fraction may rise to as much as one sixth. Given that the excess is due to gravitational lensing, it is then possible to estimate the probability that the lensing galaxy would have been detected by our survey and hence correct for the unseen component — under this hypothesis our best estimate is that one quarter to one third of the quasars in our sample are magnified into it. The allowable range is much wider than this.

To estimate the number of excess galaxies we model the galaxies as two separate populations: a group of associated galaxies which are clustered close to the quasar and a background population which are distributed randomly over the field. The cumulative distributions for the probability of finding the nng of either population within separation $r$ we take to be $g(r)$ and $f(r)$, respectively. Then the joint cumulative distribution is $g \cup f = 1 - (1 - f)(1 - g)$. Next suppose that out of a total of $N_{tot}$ fields there are $N_g$ which have associated galaxies which could have been detected. The cumulative distribution of nng separations is then

$$N(< r) = (N_{tot} - N_g)f + N_g \left( 1 - (1 - f)(1 - g) \right).$$

The distribution, $g$, of associated galaxies is unknown, however we may assume it to be centrally concentrated, i.e. $g \approx 1$ for $r \gtrsim 30$ arcsec (see Section 4.2). Then

$$N(< r) \approx (N_{tot} - N_g)f + N_g, \qquad \text{for } r > r_{nng}.$$

Finally, if we use $s$ as the radial co-ordinate instead of $r$ then $f$ is uniform and we have an expression in terms of one free parameter, $N_g$. By comparing this distribution with the observed separations we can determine a confidence range for $N_g$.

For a given value of $s$ the most likely value for $N_g$ is simply

$$N_g = \frac{N - sN_{tot}}{1 - s}.$$

If $s$ coincides with the value for one of the nng then this should be modified to read

$$N_g = \frac{N - s(N_{tot} + 1)}{1 - s}.$$

Note that these expressions can be negative. It is potentially more useful to construct a confidence range for $N_g$ which we do by the following procedure. First we pick a value of $s = s_o$ (and hence $N_o = N(< s_o)$). Then we run through a range of possible values for $N_g$. For each $N_g$ the distribution of measured values of $N$ is

$$\text{Prob}\{N < N_o\} = I(N - N_g, N_{tot} - N + 1; s_o)$$

where $I$ is the Incomplete Beta function. We reject all values of $N_g$ which make this probability less than 2.5 percent or greater than 97.5 percent.

Figure 3 shows the results of these calculations for the sample defined by $z_q > 1.0$. The solid line shows the expected value of $N_g$ as a function of $s$ while the dotted lines show the 95 percent confidence range; the lower dotted line is zero. Note that the predicted $N_g$ goes to zero at small separations because our assumption that $g \approx 1$ breaks down there. $N_g$ should tend to a constant at larger values of $s$ but the confidence interval is large and becomes unconstraining as $s$ tends to unity. The expected value of $N_g$ at $s = 0.1$ is shown in column 6 of Table 4. It is difficult to know how to combine the constraints provided by different values of $s$ as they are clearly not all independent. Column 7 of Table 4 shows the allowable range for $N_g$ at the largest value of $s$ which is less than 0.1.

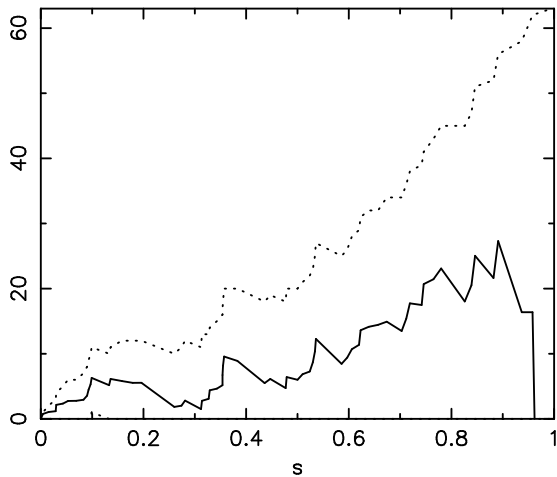

**Figure 3.** The number of excess galaxies versus $s$ for the sample, $z_q > 1.0$. The dotted lines show the 95 percent confidence range assuming a distribution of the form discussed in the text. The lower limit is zero.

The numbers in the Table need to be corrected before we can get an accurate estimate of the number of fields with an excess galaxy. Firstly we have to correct for incompleteness. We have already estimated, in Section 3.2, that about one field in thirty is missing a foreground galaxy, hidden under the quasar image. In addition there will be associated galaxies which have also been missed. It is hard to ascertain their number as we do not know the intrinsic distribution of impact parameters. The measured number density of galaxies around quasars to a magnitude limit of $m_R = 22$, shown in Table 5, increases down to approximately 3.5 arcsec then declines at smaller separations. Let us split the data up into 4 bins: 30–10 arcsec, 10–5 arcsec, 5–3.53 arcsec and < 3.53 arcsec. As a rough estimate, consistent with our earlier analysis, let us assume one missing galaxy in the inner bin and one half in the next smallest. This corresponds to zero probability of detecting a foreground galaxy at radii less than 3.53 arcsec increasing to complete detection at radii exceeding 5 arcsec. The number density of galaxies in each of the four bins is then $4.7 \times 10^{-4}$, $7.4 \times 10^{-4}$, $1.41 \times 10^{-3}$ and $8.1 \times 10^{-4}$ galaxies per square arcsecond, respectively. Now at the very least we would expect the observed number density is the inner bin to be equal to that in the next one out, which would imply at least 1.5 missing galaxies. In fact this number is very much a lower estimate because we expect to miss some associated galaxies at radii larger than 3.53 arcsec and because there is every reason to suppose that the number density of such objects should continue to rise steeply to small separations. As a conservative estimate, therefore, we suggest that the raw figure of 6.3 excess galaxies should be revised upwards to at least 9.3 (to allow for 1.5 missing foreground plus plus 1.5 missing associated galaxies).

The redshift range in which we might detect a nng does not extend all the way out to the quasar (indeed we have deliberately forced this to be the case). Given the lensing hypothesis we can predict the expected redshift distribution of lensing galaxies and hence correct for the unseen component. The lensing probability per unit redshift interval is proportional to

$$\begin{cases} \frac{[(1+z_q)^2 - (1+z)^2][(1+z)^2 - 1]}{[(1+z)^2 - 1](1+z)^{2-\delta}}, & q_0 = 0 \\ \frac{[(1+z_q)^{2.5} - (1+z)^{2.5}][(1+z)^{2.5} - 1]}{[(1+z)^{2.5} - 1](1+z)^{3.5-\delta}}, & q_0 = 0.5 \end{cases}$$

where the comoving number density of galaxies is assumed to evolve as $(1+z)^\delta$ and we assume a deceleration parameter $0 \le q_0 \le 0.5$. Taking $\delta = 0$ and $q_0 = 0.5$ gives a minimum observed fraction although it turns out that there is little difference between the predictions for different values of $q_0$. We have taken a nng luminosity of $0.4L_*$ to evaluate the fractional lens path (flp) which is accessible by our observations in each field as this is the typical luminosity of those nng with known redshifts (see Section 4.2). The results are shown in Column 8 of Table 4. If we use the flp to correct our estimate of the number of excess galaxies for the basic sample, $z_q > 1.0$, then it rises once more from 9.3 to 20.2. This is the basis for our statement that as much as one third of our quasar sample may be lensed. Using $L_*$ as a more conservative estimate of the nng luminosity gives 14.8 lensed quasars or just under one quarter of the whole sample.

We can carry out a similar sort of analysis for other samples. The subset of quasars with $M_R < -26.5$ gave 4.7 excess galaxies in 48 fields. This gets revised upwards to 7.0 to correct for incompleteness and then 15.9 to allow for the flp searched. This is about the same fraction as for the full sample. Those quasars with $z_q > 1.5$, on the other hand, give a larger signal. The numbers in this case are 6.0 excess galaxies in 46 fields rising to 8.2 and 20.0 after correcting for incompleteness and the flp. The fraction of high-redshift quasars in our sample which are lensed could be as large as 0.4–0.5.

## 4 DISCUSSION

### 4.1 Validity of the results

We pause here to reconsider the possibility that there may be a flaw in our arguments.

(i) The most obvious and serious objection would be that we started with a biased quasar sample. The quasars were selected from catalogues which have been used to study the statistics of absorption-line systems. We did not select the quasars from within this list as we observed all of our potential targets, subject to constraints on telescope time and position. However the traditional colour and morphological criteria used to select quasar candidates, may have introduced bias. For example, if only stellar images are selected as quasar candidates, a bias will be introduced against quasar images with a nearby galaxy which is within about a magnitude of the quasar magnitude. Or, if the quasar colour is reddened when seen behind a foreground galaxy, then the quasar will not be selected as part of a sample of UV-excess images. Both these effects will mean that we have underestimated the number of quasars with associated galaxies.

(ii) Is it possible that our selection criterion for faint images, or the discrimination between stars and galaxies, is altered in the presence of a nearby, bright point source? To check for this possibility we repeated our analysis for the sample of control stars defined in Section 2.2. The results are presented in Table 4 for a single magnitude cut but similar results are obtained in every case we have tried: the distribution of stars and galaxies around control stars is consistent with a Poisson distribution over the whole field.

(iii) One potential source of error would be the inclusion of galaxies which are at the quasar redshift. We circumvent this possibility by including only those fields in which the quasar redshift is greater than unity. As an alternative we also considered eliminating those fields for which the nng would have a luminosity of less than $10 L_*$ or $30 L_*$ if at the quasar redshift. This calculation, described in the Appendix of Paper 1, applies a standard k-correction but does not allow for evolution of the stellar population. The overdensities in each case are similar.

(iv) We have checked the sensitivity to detection of faint objects as a function of position on the field. For example vignetting might lower the background galaxy count relative to that under the quasar and so give artificially low values of $s$. The results of our tests, however, eliminate this possibility. When averaged over the whole sample, the mean surface density of galaxies down to $m_R = 22$ for the whole frame, but excluding the region within 30 arcsec of the quasar, is $4.26 \times 10^{-4} \mathrm{arcsec}^{-2}$, whereas the density within the central third of the field, once again avoiding the quasar, is $4.28 \times 10^{-4} \mathrm{arcsec}^{-2}$. These two values agree within the counting statistics. A similar calculation for the region *within* 30 arcsec of the quasar produces a higher value, $5.00 \times 10^{-4} \mathrm{arcsec}^{-2}$, and represents an excess of about 13 galaxies in the 63 fields with $z_q > 1.0$.

The equivalent numbers for faint stars ($m_R \geq 18$) are consistent with a uniform distribution over the whole field.

## 4.2 Properties of nearest-neighbour galaxies

The measured number counts of galaxies and stars down to $m_R = 22$, avoiding the region within 30 arcsec of the quasar in each field, is shown in Figure 4. Also shown is a theoretical distribution based on a constant comoving Schechter luminosity function

$$dN = \frac{\mathcal{L}}{L_* \Gamma(2 + \alpha)} \left( \frac{L}{L_*} \right)^\alpha \exp \left\{ -\frac{L}{L_*} \right\} d \left( \frac{L}{L_*} \right)$$

where we have taken $\alpha = -1.25$, $L_* = 10^{10} h^{-2} \mathrm{L}_\odot$ and $\mathcal{L} = 1.7 \times 10^8 h \mathrm{L}_\odot \mathrm{Mpc}^{-3}$ (the Hubble constant, $H_0 = 100 h \mathrm{\,km\,s^{-1}\,Mpc^{-1}}$), and we have included the colour and k-corrections appropriate for an Sbc galaxy (see the Appendix of Paper 1 for details). The form and normalisa-tion of the distribution agrees with observations (Colless, private communication) and with our theoretical prediction based on an unevolving galaxy population, and changes only slightly as the galaxy type and slope, $\alpha$, of the luminosity function are varied. The distribution of galaxy magnitudes is consistent with the theoretical one except at faint magnitudes, $m_R > 21.5$, where incompleteness and possible misidentification of galaxies as stars is clearly apparent.

Also shown in Figure 4 are the number countss of all nng and of those with $s < 0.1$. These show a slight bias to faint magnitudes but the numbers are so small that they are statistically indistinguishable from the whole population.

It is possible to determine neither luminosities for the nng, nor impact parameters between the nng and the line-of-sight to the quasar, without knowing the distance to the nng. We can get such information for a partial subset of all the nng if we assume that they are associated with MgII absorption lines in quasar spectra, where detected. This hypothesis has been discussed in Paper 1. There are just 10 nng which we are confident can be linked with the MgII absorption line systems. Their luminosities are consistent with being drawn from a Schechter luminosity function with $L \gtrsim 0.1 L_*$ (although the significance of this is weak), and the impact parameters extend to at least $30 h^{-1}$ kpc. Note that six of these quasars were biased with respect to the presence of MgII absorption in their spectra and are therefore omitted from the current sample. However we do not expect that this bias will affect the properties of the nng. No new quasars with known absorption redshifts have been added to the sample and so we cannot improve on these results.

If we assume that each of the nng with $s < 0.1$ has a luminosity of $0.4 L_*$, typical of the values found in Paper 1, then the set of impact parameters (for a flat Universe) is 7.9, 9.4, 10.2, 13.7, 14.3, 22.5, 23.7, 23.7, 24.4, 24.9, 31.6 and 39.3 $h^{-1}$ kpc. Of these we expect about 4–5 to be random associations. These numbers are consistent with our conclusions from Paper 1 that the impact parameters extend out to about $30 h^{-1}$ kpc.

We have checked for clustering of galaxies around the nng by doing a nearest neighbour analysis centred on the nng itself (see final entries in Table 4). The distribution of s-values is uniform (even for nng with $s < 0.1$). Note, however, that even at a redshift of 0.5 the width of our field corresponds to just 1 Mpc and so a moderately-sized group would fill the whole frame.

## 4.3 Cosmological implications

The fraction of quasars in our sample which possess an excess galaxy is of order one third, although the allowable range is large. As we have taken every precaution to eliminate galaxies which are physically associated with the quasars, we interpret this result as being due to lensing of the quasars, which would otherwise be too faint to be included in our sample.

A number of authors (e.g. Schneider 1989 and references therein) have made theoretical predictions of the expected number of quasar-galaxy associations. This depends upon two main factors, the first of which is the total mass associated with galaxies (the models assume all the mass of the galaxy is in the form of compact objects). The calculations usually model galaxies as singular isothermal spheres which have a single free parameter, the velocity dispersion. The galaxies are assumed to be distributed uniformly in space with luminosities drawn from a Schechter function as described in Section 4.2. These luminosities are related to the velocity dispersion using the Faber-Jackson relation (Faber & Jackson 1976).

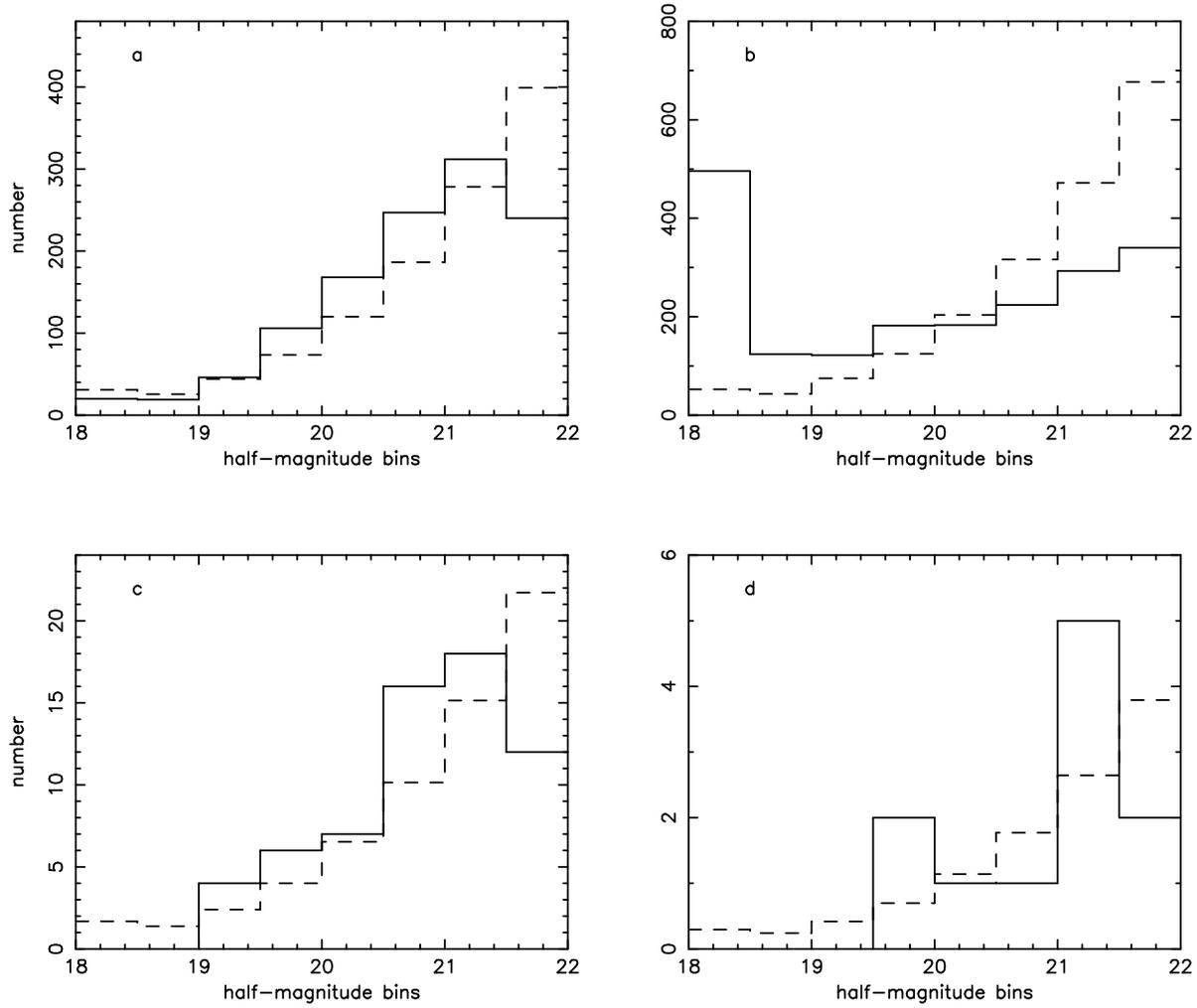

**Figure 4.**

**Figure 4.** The measured luminosity function down to a magnitude limit of 22 for: (a) all galaxies, (b) all stars, (c) nearest-neighbour galaxies for the basic sample, $z_q > 1.0$, (d) nng for the basic sample and for which $s < 0.1$. The leftmost bin in each case contains the total number of objects with $m_R < 18.5$. The dotted line shows the theoretical galaxy luminosity function for the same total number of objects.

The second input into the theoretical calculation is the intrinsic quasar luminosity function. A broken power-law (Boyle et al. 1990) is the usual description, however there has been a recent suggestion by Hawkins & Véron (1993) that the quasar luminosity function may remain steep, rather than turning over at $m \sim 19$. This would mean that theoretical predictions of the effects of lensing have been too low. Note that if a significant number of bright quasars are lensed, then the intrinsic luminosity function may well be steeper than the observed one (Webster 1992).

Schneider calculates several quantities including $q$, the fraction of quasars which have a galaxy within a given radius under the assumed lensing model, and $f$, the ratio of this number to that expected in the absence of lensing. We can use the ratio $(f-1)q/f$ to estimate the fraction of quasars which show an excess galaxy. It is difficult to translate these theoretical estimates into numbers which can be compared with the observations (as, for example, the predictions take no account of k-corrections). However, for parameters which seem representative of our sample we find a fraction 0.09; this is to be compared with our measured value of 0.15 (but with very large error bars). With more precise theoretical modeling one might hope to turn the calculations around and use the observations to constrain the mass fraction of compact objects in galaxies.

If the mass associated with luminous galaxies is mostly in the form of CDM or some other elementary particle, then only macrolensing will be important, and high amplification events will only occur when the source is near a critical line in the source plane. However, if the associated mass is in the form of compact objects, then high amplification events can occur at much larger angular radii than the critical radius, albeit with lower probabilities. Integrated over a wide area, these probabilities can become significant. There have been several observational results which suggest that microlensing of quasars might be important. Francis & Koratkar (1994) find that the differences in UV spectra between low- and high-redshift quasars are explained if 30 percent of the high-redshift sample are microlensed, and Hawkins (1993) has suggested that the long-term variability seen in the majority of quasars may be the result of microlensing by brown dwarfs. The possible detection of MACHOs in the halo of our galaxy (Alcock et al. (1993; Aubourg et al. 1993) strengthens the possibility that microlensing might contribute significantly to quasar-galaxy associations. If many of the associations are due to microlensing by compact objects in extended regions with optical depths to lensing of $\lesssim 0.1$, then this would explain in a natural way why they are much more common than multiple images. We note in passing that the large number of wide-separation binary quasars, which are unverified as lenses, might be a link in this sequence. This idea has been explored by Bartelmann and Schneider (1991).

If microlensing is important, then differential amplification effects, as discussed by Canizares (1981), might come into play. Thus, for example, the spectra of the microlensed quasars might show an enhanced blue continuum, or weaker emission line equivalent widths. We have tried analysing the radio-loud quasars separately (*i.e.* all those flagged with an R in Column 2 of Table 1). The result, shown in Table 4, is more significant than for the sample as a whole. This is especially surprising given the smaller number of fields. It is perplexing that the associations seem more significant for the radio-loud quasars, since it is generally thought that the radio emission region is substantially larger than the optical region, and is therefore unlikely to be microlensed. In principle amplification (or the lack of it) can be used to constrain the mass of putative microlenses.

## 5 CONCLUSIONS

In this paper we search for galaxies lying close to the line-of-sight to a sample of 63 quasars. The quasar-nearest neighbour galaxy separation is described by the 's-parameter' which is designed so as to give a uniform distribution for a random galaxy positions on the field—we find a significant bias to low values of $s$, i.e. small separations. We have looked at many possible biases which might cause this result but can find no plausible explanation other than gravitational lensing which may magnify faint quasars into our sample.

The fraction of quasars which possess a nearby, excess galaxy is about one sixth but rises to one third once we correct for the fraction of lensing galaxies which would be too faint to be detected. This result has implications both for the intrinsic quasar luminosity function and for the mass density of matter locked-up in compact objects capable of producing the observed microlensing. We note, however, that the allowable error range for the fraction of lensed quasars is large.

Future work must involve obtaining as many spectra as possible for the galaxies near to the quasars in our sample. Redshifts are essential both to test the lensing hypothesis and to determine the properties of the excess galaxies. We would also be able to investigate more accurately the link between galaxies and MgII absorption lines discussed in Paper 1.

## Acknowledgments

PAT and RLW would like to thank CITA, and MJD would like to thank the Physics Dept., Laval Université, for support during the early stages of this project. MJD thanks MESS (Québec Government) for travel support and the SERC for support while a visitor at the Astronomy Centre, Sussex. The data were obtained at l'Observatoire du Mont Mégantic, Qu'ebec which is supported by grants from NSERC (Canadian Government) and FCAR (Québec Government). We are grateful to Bernard Malenfaut and Ghislain Turcotte for their assistance at the telescope, and to Mme. Brault for her excellent cuisine which kept us fortified. Part of the data analysis was carried out on the STARLINK minor node at Sussex using the IRAF software supported by the NOAO.

# REFERENCES

Alcock C. et al. , 1993, Nat, 365, 621

Aubourg E. et al. , 1993, Nat, 365, 623

Bartelmann M., Schneider P., 1991, A&A, 239, 113

Canizares C. R., 1981, Nat, 291, 620

Christian C. A., Crabtree D., Waddell P., 1987, ApJ, 312, 45

Crampton D., McClure R. D., Fletcher J. M., 1992, ApJ, 392, 23

Drinkwater M. J., Webster R. L., Thomas P. A., 1993, AJ, 106, 848 (Paper 1)

Drinkwater M. J., Webster R. L., Thomas P. A., Millar D., 1992, Publ. astr. Soc. Aust., 10, 8

Ellingson E., Green R. F., Yee H. K. C., 1991, ApJ, 378, 476

Faber S. M., Jackson R. E., 1976, ApJ, 204, 666

Foltz C. B., Weyman R. J., Peterson B. M., Sun L., Malkan M. A., Chaffee F. H., 1986, ApJ, 307, 50

Francis P. J., Koratkar A. P., 1994, Preprint

Fugmann W., 1988, A&A, 204, 73

Fugmann W., 1989, A&A, 222, 45

Hawkins M. R. S., 1993, Nat, 366, 242

Hawkins M. R. S., Véron P., 1993, MNRAS, 260, 202

Hewett P. C., Harding M. E., Webster R. L., 1992, in Kayser R., Schramm T., Nieser L., eds., Gravitational Lenses, Hamburg. Springer, Berlin, p. 209

Hewitt A., Burbidge G., 1989, ApJS, 69, 1

Hintzen P., Romanishin W., Valdes F., 1991, ApJ, 366, 7

Sargent W. L. W., Steidel C. C., Boksenberg A., 1988, ApJ, 334, 2

Schneider P., 1989, A&A, 221, 221

Smith E. P., Heckman T. M., 1990, ApJ, 348, 38

Tyson J. A., 1986, AJ, 92, 691

Tytler D., Boksenberg A., Sargent W. L. W., Young P. J., Kunth D., 1987, ApJS, 64, 66

Valdes F., 1982, FOCAS User's Manual. Kitt Peak National Observatory, Tucson

Webster R. L., 1992, in Crampton D., ed., The Space Distribution of Quasars. A.S.P. Conference Series Vol. 21, p. 160

Webster R. L., Hewett P. C., Harding M. E., Wegner G. A., 1988, Nat, 336, 358

Wegner G., McMahan R. K., 1987, AJ, 94, 1271

Wegner G., McMahan R. K., 1988, AJ, 96, 1933

Wegner G., Swanson S. R., 1990a, AJ, 99, 330

Wegner G., Swanson S. R., 1990b, AJ, 100, 1274

Weymann R. J., Williams R. E., Peterson B. M. , Turnsheck D. A., 1979, ApJ, 234, 3

Yee H. K. C., Green R. F., 1987, ApJ, 319, 28

Yee H. K. C., Filippenko A. V. , Tang D., 1992, in Kayser R., Schramm T., Nieser L., eds., Gravitational Lenses, Hamburg. Springer, Berlin, p. 83

Young P., Sargent W. L. W., Boksenberg A., 1982, ApJS, 48, 45